\documentclass[twocolumn, 10pt]{IEEEtran}
\usepackage{amsmath,amssymb,amsthm}
\usepackage{bm}
\usepackage{url}
\usepackage{tikz}
\usepackage[caption=false,font=footnotesize]{subfig}

\newcommand{\Var}{\mathrm{Var}}
\newcommand{\Cov}{\mathrm{Cov}}
\newcommand{\Cor}{\mathrm{Cor}}
\newcommand{\tr}{\mathrm{tr}}

\newtheorem{theorem}{Theorem}
\newtheorem{lemma}[theorem]{Lemma}
\newtheorem{corollary}[theorem]{Corollary}
\theoremstyle{definition}
\newtheorem{definition}[theorem]{Definition}
\newtheorem{example}[theorem]{Example}
\theoremstyle{remark}
\newtheorem{remark}{Remark}

\urldef\rmmail\url{mori@is.titech.ac.jp}

\title{Loop Calculus for Non-Binary Alphabets using Concepts from Information Geometry}
\author{Ryuhei~Mori~\IEEEmembership{Member,~IEEE}%
\thanks{This paper was presented in part at the 2012 IEICE Symposium on Information Theory and its Application.}%
\thanks{This work was supported by MEXT KAKENHI Grant Number 24106008.}%
\thanks{R. Mori is with the Department of Mathematical and Computing Science, Graduate School of Information Science and Engineering,
Tokyo Institute of Technology, Shibaura, Minato-ku, Tokyo, 108-0023 Japan
(e-mail: \rmmail).}%
}

\begin{document}
\maketitle
\begin{abstract}
The Bethe approximation is a well-known approximation of the partition function used in statistical physics.
Recently, an equality relating the partition function and its Bethe approximation was obtained for graphical models with binary variables by Chertkov and Chernyak.
In this equality, the multiplicative error in the Bethe approximation is represented as a weighted sum over all generalized loops
 in the graphical model.
In this paper, the equality is generalized to graphical models with non-binary alphabet
using concepts from information geometry.
\end{abstract}

\section{Introduction}
\IEEEPARstart{C}{omputing} the partition function
is one of the central problem in statistical physics, information theory, machine learning and computer science.
While the exact computation of the partition function is generally $\mathsf{\#P}$-hard,
the Bethe approximation provides an easily computable estimate whose accuracy is quite good for many problems~\cite{bethe1935statistical}, \cite{mezard2009ipa}.
The Bethe approximation is traditionally defined by a heuristic method called the cluster variation method (CVM)~\cite{pelizzola2005cluster}.
However, it is generally difficult to give theoretical guarantees on the Bethe approximation from the idea of the CVM.
Recently, Chertkov and Chernyak showed an equality relating the partition function and its Bethe approximation for graphical models with the binary alphabet, which is
\begin{equation}
Z(G) = Z_{\mathrm{Bethe}}(G)\left(1+\sum_{\gamma\in\mathcal{G}}\mathcal{K}(\gamma)\right)
\label{eq:LC}
\end{equation}
where $Z(G)$ is the partition function, $Z_{\mathrm{Bethe}}(G)$ is its Bethe approximation, $\mathcal{G}$ is a set of subsets of edges called generalized loops and $\mathcal{K}(\gamma)$ is a (possibly negative) weight of a generalized loop $\gamma\in\mathcal{G}$~\cite{chertkov2006loop}.
This equality means that the error in the Bethe approximation can be expressed as the weighted sum over all generalized loops 
in the graphical model.
In contrast to other well-known loop (or diagram) series expansions of the Gibbs free energy and the entropy functionals~\cite{georges1991expand},
\cite{sessak2009small}, in Chertkov and Chernyak's loop calculus~\eqref{eq:LC},
the error in the Bethe approximation is represented by the loop series.
From this property,~\eqref{eq:LC}
is useful 
not only for improvement of the Bethe approximation~\cite{chertkov2006lp}, \cite{GomezMooijKappen_JMLR_07}
but also for bounding the error in the Bethe approximation~\cite{chandrasekaran2011counting}, \cite{macris2012beyond}
although the exact computation of the summation for generalized loops is still $\mathsf{\#P}$-hard in general.
The equality~\eqref{eq:LC} is also generalized for more general CVM approximation~\cite{zhou2012region}.
Similar expansion of the error in the Bethe approximation is recently known in~\cite{conf/uai/WellingGI12}.

While~\eqref{eq:LC} was derived based on some properties of fixed point of belief propagation in~\cite{sudderth2008loop},
Chertkov and Chernyak derived~\eqref{eq:LC} based on a general equality, called holographic transformation,
with particular constraints that requires zero weights for non-loop structures
in~\cite{chertkov2006loop}, \cite{1742-5468-2006-06-P06009}, \cite{chernyak2007loop}.
Since the Bethe approximation naturally appears from the constraints, it gives a new characterization of the Bethe approximation.
Furthermore,  in~\cite{chernyak2007loop},~\eqref{eq:LC} is generalized to graphical models with non-binary alphabets in a recursive way.
However, the representation of the equation is less explicit and there sometimes exist difficulties in the recursive method.
Based on the derivation of Chernyak and Chertkov in~\cite{chernyak2007loop} and concepts from information geometry,
this paper derives~\eqref{eq:LC} in an explicit form for general graphical models with non-binary alphabets.
Our equations cover all equations obtained by Chertkov and Chernyak's idea.
This result is useful for the Bethe approximation for graphical models with non-binary alphabets and 
also for region-based approximation for binary graphical models~\cite{zhou2012region}.

This paper is organized as follows.
In Section~\ref{sec:factor}, the factor graph model and notations used in this paper are defined.
In Section~\ref{sec:Bethe}, the Bethe approximation and belief propagation are defined.
In Section~\ref{sec:hol}, generalizations of the loop calculus to non-binary alphabets are shown, which are the main results of this paper.
In the section, we use tools of information geometry shown in Appendix~\ref{apx:ef}.
In Section~\ref{sec:simple}, it is shown that the weights of simple generalized loops for non-binary alphabets can be represented
 as trace of product of matrices.
In Section~\ref{sec:lcont}, the loop calculus is generalized to continuous alphabets, which is originally obtained by
Xiao and Zhou~\cite{xiao2011partition}.
Furthermore, it is simplified for the Gaussian model.
In Section~\ref{sec:color}, simple examples of improvement of the approximation using loop calculus are shown by numerical calculations
for the weighted coloring problem.

\section{Factor graph and preliminaries}\label{sec:factor}
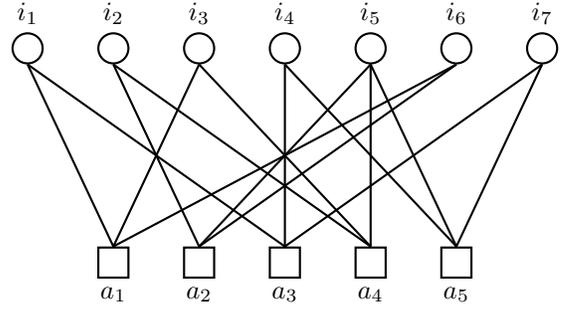
\begin{figure}
\centering
\begin{tikzpicture}
[yshift=20pt, scale=0.57, inner sep=0mm, C/.style={minimum size=4mm,circle,draw=black,thick},
S/.style={minimum size=4mm,rectangle,draw=black,thick}, label distance=1mm]
\node (a1) at (0,0) [S,label=below:$a_1$] {};
\node (a2) at (2,0) [S,label=below:$a_2$] {};
\node (a3) at (4,0) [S,label=below:$a_3$] {};
\node (a4) at (6,0) [S,label=below:$a_4$] {};
\node (a5) at (8,0) [S,label=below:$a_5$] {};
\node (1) at (-2,5.0) [C,label=above:$i_1$] {};
\node (2) at (0,5.0) [C,label=above:$i_2$] {};
\node (3) at (2,5.0) [C,label=above:$i_3$] {};
\node (4) at (4,5.0) [C,label=above:$i_4$] {};
\node (5) at (6,5.0) [C,label=above:$i_5$] {};
\node (6) at (8,5.0) [C,label=above:$i_6$] {};
\node (7) at (10,5.0) [C,label=above:$i_7$] {};
\draw (a1.north) to (1.south) [thick];
\draw (a1.north) to (3.south) [thick];
\draw (a1.north) to (6.south) [thick];
\draw (a2.north) to (2.south) [thick];
\draw (a2.north) to (5.south) [thick];
\draw (a2.north) to (6.south) [thick];
\draw (a3.north) to (1.south) [thick];
\draw (a3.north) to (4.south) [thick];
\draw (a3.north) to (7.south) [thick];
\draw (a4.north) to (2.south) [thick];
\draw (a4.north) to (3.south) [thick];
\draw (a4.north) to (5.south) [thick];
\draw (a5.north) to (4.south) [thick];
\draw (a5.north) to (5.south) [thick];
\draw (a5.north) to (7.south) [thick];
\end{tikzpicture}
\caption{An example of factor graph. Variable nodes and factor nodes are described by circles and squares, respectively.
The set of variable nodes, the set of factor nodes and the set of edges are
$V=\{i_1, i_2, i_3, i_4, i_5, i_6, i_7\}$, $F=\{a_1,a_2,a_3,a_4,a_5\}$ and
$E=\{(i_1,a_1),(i_1,a_3), (i_2,a_2), (i_2,a_4),\allowbreak (i_3,a_1), (i_3,a_4),\allowbreak (i_4,a_3), (i_4,a_5),\allowbreak
(i_5,a_2), (i_5,a_4), (i_5,a_5),\allowbreak
(i_6,a_1), (i_6,a_2),\allowbreak (i_7,a_3), (i_7,a_5)\}$, respectively.}
\label{fig:factor}
\end{figure}

In this paper, we deal with a general graphical model called a factor graph~\cite{RiU05/LTHC}.
Let $V$ and $F$ be a set of variable nodes and a set of factor nodes, respectively.
Let $E\subseteq V\times F$ be a set of edges.
Let $\partial i\subseteq F$ and $\partial a\subseteq V$ be neighborhoods of $i\in V$ and $a\in F$, respectively.
Let $d_i$ and $d_a$ be a degree of variable node $i\in V$ and a degree of factor node $a\in F$, respectively.
For each variable node $i\in V$, there is a corresponding variable $x_i$ taking a value on the alphabet set $\mathcal{X}_i$.
The alphabet sets are assumed to be finite unless otherwise stated. For the simplicity, we assume that the alphabet set is common for all $i\in V$
and hence is denoted by $\mathcal{X}$.
Let $q$ be the cardinality of the alphabet $\mathcal{X}$.
For $C\subseteq V$, $\bm{x}_C$ denotes $(x_i)_{i\in C}$.
For each variable node $i\in V$ and a factor node $a\in F$, there are corresponding functions
$h_i\colon \mathcal{X}\to\mathbb{R}_{> 0}$ and $f_a\colon \mathcal{X}^{d_a}\to\mathbb{R}_{\ge 0}$, respectively.
Here, it is assumed that $d_a\ge 2$ for all $a\in F$.
Let $N$ be the number of variable nodes in a factor graph.
Then, the probability measure on $\mathcal{X}^N$ defined by the factor graph $G=(V, F, E, (h_i)_{i\in V}, (f_a)_{a\in F})$ is
\begin{equation*}
p(\bm{x};G):= \frac1{Z(G)} \prod_{a\in F} f_a(\bm{x}_{\partial a}) \prod_{i\in V} h_i(x_i)
\end{equation*}
where $Z(G)$ is a constant for the normalization defined by
\begin{equation*}
Z(G) := \sum_{\bm{x}\in\mathcal{X}^N} \prod_{a\in F} f_a(\bm{x}_{\partial a}) \prod_{i\in V} h_i(x_i).
\end{equation*}
The constant $Z(G)$ is called a partition function.
Historically, $Z(G)$ has been defined for exponential families. Hence, $Z(G)$ can be regarded as a function of natural parameters (see also Appendix~\ref{apx:ef}).
This is the reason why $Z(G)$ is called a partition \textit{function}.
Generally the computation of the partition function is $\mathsf{\#P}$-hard.
Hence, the efficient accurate approximation is a worthwhile goal.
An example of a factor graph is shown in Fig.~\ref{fig:factor}.
If all variable nodes have degree 2, a factor graph is also called a normal factor graph~\cite{5695119}, \cite{forney2011partition}.
General factor graph can be transformed to a normal factor graph with the same partition function
by replacing edges by degree-2 variable nodes and by replacing variable nodes by the equality constraints.

Let $\mathcal{S}_a\subseteq\mathcal{X}^{d_a}$ be the support of $f_a$, i.e., $\mathcal{S}_a:=\{\bm{x}_{\partial a}\in\mathcal{X}^{d_a}\mid f_a(\bm{x}_{\partial a})>0\}$
for all $a\in F$.
Assume that $\mathcal{S}_a\cap \{\bm{x}_{\partial a}\in\mathcal{X}^{d_a}\mid x_i = z\}\ne\varnothing$ for any $(i,a)\in E$ and $z\in\mathcal{X}$.
Let $\mathcal{S}\subseteq\mathcal{X}^N$ be the support of $p$, i.e., 
$\mathcal{S}:=\bigcap_{a\in F}\{\bm{x}\in\mathcal{X}^N\mid \bm{x}_{\partial a}\in\mathcal{S}_a\}$.
Let $\mathcal{P}(\Omega)$ be the set of probability measures on a set $\Omega$.
Let $\langle \cdot\rangle_{p'}$ be the expectation with respect to a probability measure $p'$.
Let $\delta(x,y)$ be a function taking 1 if $x=y$ and 0 if $x\ne y$.
Let $|\mathcal{A}|$ be the cardinality of a set $\mathcal{A}$.
For any $x,z\in\mathcal{A}$, let $L_{x,z}$ be the $(x,z)$-element of a matrix $L$ indexed by an element of a set $\mathcal{A}$.
Let $L^t$ be the transpose of a matrix $L$.

\section{Bethe approximation and belief propagation}\label{sec:Bethe}
In this section, the Bethe approximation is defined for an approximation of the partition function of a factor graph.
The Bethe approximation is defined based on a variational representation of the partition function.

\begin{definition}[Gibbs free energy]
For $p'\in\mathcal{P}(\mathcal{S})$, the Gibbs free energy of a factor graph $G$ is defined as
\begin{equation*}
\mathcal{F}_{\mathrm{Gibbs}}(p') := \mathcal{U}_{\mathrm{Gibbs}}(p') - \mathcal{H}_{\mathrm{Gibbs}}(p')
\end{equation*}
where
\begin{align*}
\mathcal{U}_{\mathrm{Gibbs}}(p') &:= 
-\sum_{a\in F}\sum_{\bm{x}\in\mathcal{S}} p'(\bm{x})\log f_a(\bm{x}_{\partial a})\\
&\qquad -\sum_{i\in V}\sum_{\bm{x}\in\mathcal{S}} p'(\bm{x})\log h_i(x_i)\\
\mathcal{H}_{\mathrm{Gibbs}}(p') &:= -\sum_{\bm{x}\in\mathcal{S}} p'(\bm{x})\log p'(\bm{x}).
\end{align*}
Here, $\mathcal{U}_{\mathrm{Gibbs}}$ and $\mathcal{H}_{\mathrm{Gibbs}}$ are called the Gibbs average energy and the Gibbs entropy, respectively.
\end{definition}

Then, it holds $-\log Z(G) = \min_{p'}\mathcal{F}_{\mathrm{Gibbs}}(p')$.
The minimum is achieved by $p'=p$.
Instead of approximating the partition function directly, we consider an approximation for the Gibbs free energy.
Since the domain $\mathcal{P}(\mathcal{S})$ of the Gibbs free energy generally needs exponentially many variables and inequalities,
a polytope consisting of marginal distributions is considered for approximations.

\begin{definition}[Marginal polytope]
The marginal polytope $\mathcal{M}(G)\subseteq\mathcal{P}(\mathcal{X})^N\times\prod_{a\in F}\mathcal{P}(\mathcal{S}_a)$ is defined by
\begin{align*}
&\mathcal{M}(G):=\Bigl\{((p'_i\in\mathcal{P}(\mathcal{X}))_{i\in V},(p'_a\in\mathcal{P}(\mathcal{S}_a))_{a\in F})\mid\\
&\qquad \exists p'\in \mathcal{P}(\mathcal{S}),\, p'_i(z_i)=\sum_{\bm{x}\in\mathcal{S}, x_i=z_i}p'(\bm{x}),\,\forall i\in V, \forall z_i\in\mathcal{X},\\
&\qquad p'_a(\bm{z}_{\partial a})=\sum_{\bm{x}\in\mathcal{S}, \bm{x}_{\partial a}=\bm{z}_{\partial a}} p'(\bm{x}),\,\forall a\in F, \forall \bm{z}_{\partial a}\in\mathcal{X}^{d_a}\Bigr\}.
\end{align*}
\end{definition}
In general, the marginal polytope still needs exponentially many inequalities to 
guarantee the existence of a consistent global distribution $p' \in \mathcal{P}(\mathcal{X}^N)$.
For reducing the number of variables and inequalities in the representation,
a set of the marginal distributions which only satisfies local constraints is considered.

\begin{definition}[Local marginal polytope]
The local marginal polytope $\mathcal{L}(G)\subseteq\mathcal{P}(\mathcal{X})^N\times\prod_{a\in F}\mathcal{P}(\mathcal{S}_a)$ is defined by
\begin{align*}
\mathcal{L}(G)&:=\Bigl\{((b_i\in\mathcal{P}(\mathcal{X}))_{i\in V},(b_a\in\mathcal{P}(\mathcal{S}_a))_{a\in F})\mid\\
& b_a\in \mathcal{P}(\mathcal{S}_a),\,\forall a\in F,\\
& b_i(z_i)=\sum_{\bm{x}_{\partial a\setminus\{i\}, x_i=z_i}} b_a(\bm{x}_{\partial a}),
\,\forall (i,a)\in E, \forall z_i\in\mathcal{X}\Bigr\}.
\end{align*}
\end{definition}
Obviously, 
the local marginal polytope $\mathcal{L}(G)$ is a superset of the marginal polytope $\mathcal{M}(G)$, i.e., $\mathcal{M}(G)\subseteq\mathcal{L}(G)$.
If a factor graph $G$ is cycle-free, it holds $\mathcal{L}(G)=\mathcal{M}(G)$ since one can explicitly construct
a global distribution $p'\in\mathcal{P}(\mathcal{S})$ consistent with the marginal distributions
$((b_i\in\mathcal{P}(\mathcal{X}))_{i\in V}, (b_a\in\mathcal{P}(\mathcal{S}_a))_{a\in F})\in\mathcal{L}(G)$.
\begin{lemma}[\cite{wainwright2008graphical}]
If $G$ is cycle-free, $\mathcal{L}(G) = \mathcal{M}(G)$.
\end{lemma}
\begin{IEEEproof}
Since $G$ is cycle-free, $\mathcal{L}(G)$ has at least $q$ vertices.
It is sufficient to show that interior points of $\mathcal{L}(G)$ are included in $\mathcal{M}(G)$.
Assume
$((b_i\in\mathcal{P}(\mathcal{X}))_{i\in V}, (b_a\in\mathcal{P}(\mathcal{S}_a))_{a\in F})\in\mathcal{L}(G)$
satisfies $b_a(\bm{x}_{\partial a})>0$ for all $a\in F$ and $\bm{x}_{\partial a}\in\mathcal{S}_a$.
Then, we will show that
\begin{equation}
p'(\bm{x})=
\prod_{a\in F}\frac{b_a(\bm{x}_{\partial a})}{\prod_{i\in\partial a}b_i(x_i)}
\prod_{i\in V}b_i(x_i),
\hspace{2em} \bm{x}\in\mathcal{X}^N
\label{eq:tree}
\end{equation}
is a valid distribution whose support is $\mathcal{S}$, and
 is consistent with the local marginals $((b_i)_{i\in V}, (b_a)_{a\in F})$.
It is obvious that $p'$ is non-negative and $p'(\bm{x})=0$ for $\bm{x}\notin\mathcal{S}$.
One can also confirm that~\eqref{eq:tree} is normalized, i.e., $\sum_{\bm{x}\in\mathcal{S}} p'(\bm{x})=1$ as follows.
First,~\eqref{eq:tree} is expanded as
\begin{align*}
p'(\bm{x})&=
\prod_{a\in F}\left(1+\frac{b_a(\bm{x}_{\partial a})-\prod_{i\in\partial a}b_i(x_i)}{\prod_{i\in\partial a}b_i(x_i)}\right)
\prod_{i\in V}b_i(x_i)\\
&=
\sum_{F'\subseteq F}\prod_{a\in F'}\frac{b_a(\bm{x}_{\partial a})-\prod_{i\in\partial a}b_i(x_i)}{\prod_{i\in\partial a}b_i(x_i)}
\prod_{i\in V}b_i(x_i).
\end{align*}
Since a graph $G$ is assumed to be cycle-free, for any non-empty $F'\subseteq F$, there exist $a_{F'}\in F'$ and $i_{F'}\in\partial a_{F'}$
such that $(\partial a_{F'}\setminus\{i_{F'}\})\cap\partial a=\varnothing$ for all $a\in F'\setminus\{a_{F'}\}$.
Hence,
$\sum_{\bm{x}\in\mathcal{X}^N}p'(\bm{x})$ is equal to
\begin{align*}
&\sum_{F'\subseteq F}\sum_{\bm{x}\in\mathcal{X}^N}\prod_{a\in F'}\frac{b_a(\bm{x}_{\partial a})-\prod_{i\in\partial a}b_i(x_i)}{\prod_{i\in\partial a}b_i(x_i)}
\prod_{i\in V}b_i(x_i)\\
&=
1+\sum_{F'\subseteq F, F'\ne\varnothing}\sum_{\bm{x}\in\mathcal{X}^N}
\frac{b_{a_{F'}}(\bm{x}_{\partial a_{F'}})-\prod_{i\in\partial a_{F'}}b_i(x_i)}{b_{i_{F'}}(x_{i_{F'}})}\\
&\cdot\prod_{a\in F'\setminus\{a_{F'}\}}\frac{b_a(\bm{x}_{\partial a})-\prod_{i\in\partial a}b_i(x_i)}{\prod_{i\in\partial a}b_i(x_i)}
\prod_{i\in V\setminus (\partial a_{F'}\setminus \{i_{F'}\})}b_i(x_i)\\
&=1.
\end{align*}
The last equality holds since
 $\sum_{\bm{x}_{\partial a_{F'}\setminus \{i_{F'}\}}} [b_{a_{F'}}(\bm{x}_{\partial a_{F'}})-\prod_{i\in\partial a_{F'}}b_i(x_i)]=0$
holds from the definition of the local marginal polytope.
Hence, $p'\in\mathcal{P}(\mathcal{S})$.
In a similar way, it can be also shown that $((b_i)_{i\in V},(b_a)_{a\in F})$ are marginal distributions of $p'$,
 which concludes $((b_i)_{i\in V},(b_a)_{a\in F})\in\mathcal{M}(G)$ and $\mathcal{L}(G)=\mathcal{M}(G)$ for a cycle-free factor graph $G$.
\end{IEEEproof}
For a cycle-free factor graph $G$, the global distribution~\eqref{eq:tree} is a unique consistent distribution
 among all distributions on $\mathcal{S}$
which can be factorized into the form
$q(\bm{x})=\prod_{a\in F} f'_a(\bm{x}_{\partial a})$
for some $(f'_a\colon \mathcal{S}_a\to \mathbb{R}_{> 0})_{a\in F}$
since $((b_i)_{i\in V}, (b_a)_{a\in F})$ can be regarded as the expectation parameters for an appropriate exponential family~\cite{amari2000methods}.
On the other hand, the local marginal polytope is strictly larger than the marginal polytope if a factor graph includes a cycle.
For example, let $\mathcal{X}=\{0,1\}$, $V=\{i_1,i_2,i_3\}$, $F=\{a_{12}, a_{23}, a_{31}\}$,
 $E=\{(i_1, a_{12}), (i_1, a_{31}), (i_2, a_{12}), (i_2, a_{23}), (i_3, a_{23}), (i_3, a_{31})\}$ and $\mathcal{S}_a=\{0,1\}^2$ for all $a\in F$.
Let us consider $((b_i)_{i\in V}, (b_a)_{a\in F})$ defined by $b_i(0)=b_i(1)=1/2$ for all $i\in V$,
 and $b_a(0,1)=b_a(1,0)=1/2$ and $b_a(0,0)=b_a(1,1)=0$ for all $a\in F$.
Then, it satisfies all of the local constraints, and hence is an element of the local marginal polytope.
However, there is no global distribution consistent with $((b_i)_{i\in V}, (b_a)_{a\in F})$
since in any assignment on the three binary variables, at least one of the three pairs must take the same values.

Since it holds $\mathcal{M}(G)\subsetneq\mathcal{L}(G)$ in general, each element
$((b_i)_{i\in V}, (b_a)_{a\in F})$ 
in $\mathcal{L}(G)$ is called pseudo-marginals.
There are only few known exceptions of graphical model $G$ which includes cycles but $\mathcal{L}(G)=\mathcal{M}(G)$.
The most popular example would be perfect matching on the complete bipartite graph, 
for which the local marginal polytope coincides with the marginal polytope although the corresponding factor graph includes many cycles.
This fact is known as Birkhoff-von Neumann theorem,
 which states that the set of doubly stochastic matrices is equal to the convex hull of the set of the permutation matrices~\cite{vontobel2013bethe}.
Note that it is recently shown that for perfect matching on the (non-bipartite) complete graph, the marginal polytope needs exponentially many
inequalities~\cite{rothvoss2013matching}.

The Bethe free energy is an approximation of the Gibbs free energy,
which can be understood
from the above observations of the local marginal polytope and exactness for cycle-free factor graphs,

\begin{definition}[Bethe free energy~\cite{bethe1935statistical}, \cite{peierls1936statistical}]
The Bethe free energy is defined for pseudo-marginals
$((b_i)_{i\in V}, (b_a)_{a\in F})\in \mathcal{L}(G)$ as
\begin{align*}
&\mathcal{F}_{\mathrm{Bethe}}((b_i)_{i\in V}, (b_a)_{a\in F})\\
 &\quad := \mathcal{U}_{\mathrm{Bethe}}((b_i)_{i\in V}, (b_a)_{a\in F}) - \mathcal{H}_{\mathrm{Bethe}}((b_i)_{i\in V}, (b_a)_{a\in F})
\end{align*}
where
\begin{align*}
&\mathcal{U}_{\mathrm{Bethe}}((b_i)_{i\in V}, (b_a)_{a\in F})\\
&\quad := -\sum_{a\in F}\sum_{\bm{x}_{\partial a}\in\mathcal{S}_a} b_a(\bm{x}_{\partial a})\log f_a(\bm{x}_{\partial a})\\
&\qquad -\sum_{i\in V}\sum_{x_i\in\mathcal{X}} b_i(x_i)\log h_i(x_i)\\
&\mathcal{H}_{\mathrm{Bethe}}((b_i)_{i\in V}, (b_a)_{a\in F})\\
&\quad := -\sum_{a\in F}\sum_{\bm{x}_{\partial a}\in\mathcal{S}_a} b_a(\bm{x}_{\partial a})\log b_a(\bm{x}_{\partial a})\\
&\qquad +\sum_{i\in V}(d_i-1) \sum_{x_i\in\mathcal{X}}b_i(x_i)\log b_i(x_i).
\end{align*}
\end{definition}

Since for a cycle-free factor graph $G$, it holds $\mathcal{L}(G)=\mathcal{M}(G)$ and \eqref{eq:tree} is the unique consistent global distribution,
the minimization of the Gibbs free energy is equivalent to the minimization of the Bethe free energy.
From this property, the minimum of the Bethe free energy would be considered also for factor graphs with cycles as an approximation
of the minimum of the Gibbs free energy.
The Bethe free energy is often explained by the CVM~\cite{pelizzola2005cluster}.
There are also other characterizations by the Plefka expansion~\cite{georges1991expand}, \cite{0305-4470-15-6-035}, the method of graph covers~\cite{6570731}
and the loop calculus~\cite{chernyak2007loop}, which is the main topic of this paper.
Since the minimum of the Gibbs free energy is $-\log Z(G)$, the minimum of the Bethe free energy is regarded as an approximation for $-\log Z(G)$.

\begin{definition}[Bethe approximation]
The Bethe approximation for the partition function of the factor graph $G$ is defined as
\begin{align*}
&Z_{\mathrm{Bethe}}(G)\\
&:=
\exp\left\{-\min_{((b_i)_{i\in V}, (b_a)_{a\in F})\in\mathcal{L}(G)} \mathcal{F}_{\mathrm{Bethe}}((b_i)_{i\in V}, (b_a)_{a\in F})\right\}.
\end{align*}
The Bethe approximation $Z_{\mathrm{Bethe}}((b_i)_{i\in V}, (b_a)_{a\in F})$
at pseudo-marginals $((b_i)_{i\in V}, (b_a)_{a\in F})\in\mathcal{L}(G)$ is defined as
\begin{align*}
\exp\left\{-\mathcal{F}_{\mathrm{Bethe}}((b_i)_{i\in V}, (b_a)_{a\in F})\right\}.
\end{align*}
\end{definition}

When the factor graph $G$ is cycle-free, the Bethe free energy is convex
and has the unique minimum which is exactly $-\log Z(G)$
since the Bethe free energy is essentially equivalent to the Gibbs free energy due to the representation~\eqref{eq:tree} of the unique consistent global distribution.
Since the Bethe entropy is generally neither convex nor concave, it is difficult to solve the minimization problem of the Bethe free energy.
On the other hand, one can consider the stationary condition for the Lagrangian of the minimization problem,
which is a necessary condition of the minimum of the Bethe free energy.
The stationary condition for the Lagrangian of the minimization problem can be shown by using internal variables $((m_{i\to a}, m_{a\to i})\in\mathcal{P}(\mathcal{X})^2)_{(i,a)\in E}$ as follows.

\begin{lemma}[Stationary condition of the Bethe free energy~\cite{yedidia2005constructing}]\label{lem:bstat}
A pseudo-marginal $((b_i)_{i\in V}, (b_a)_{a\in F})\in \mathcal{L}(G)$ is a stationary point of the Bethe free energy
if and only if there exists a representation
\begin{align}
b_a(\bm{x}_{\partial a}) &= \frac1{Z_a((m_{i\to a})_{i\in\partial a})} f(\bm{x}_{\partial a})\prod_{i\in\partial a} m_{i\to a}(x_i)\nonumber\\
b_i(x_i) &= \frac1{Z_i((m_{a\to i})_{a\in\partial i})} h_i(x_i)\prod_{a\in\partial i} m_{a\to i}(x_i)\label{eq:pseudo}\\
&= \frac1{Z_{i,a}(m_{i\to a}, m_{a\to i})} m_{i\to a}(x_i) m_{a\to i}(x_i),\, \forall a\in\partial i\nonumber
\end{align}
where
\begin{align*}
Z_a((m_{i\to a})_{i\in\partial a}) &:= \sum_{\bm{x}_{\partial a}\in\mathcal{S}_a} f(\bm{x}_{\partial a})\prod_{i\in\partial a} m_{i\to a}(x_i)\\
Z_i((m_{a\to i})_{a\in\partial i}) &:= \sum_{x_i\in\mathcal{X}} h_i(x_i)\prod_{a\in\partial i} m_{a\to i}(x_i)\\
Z_{i,a}(m_{i\to a}, m_{a\to i}) &:= \sum_{x_i\in\mathcal{X}} m_{i\to a}(x_i) m_{a\to i}(x_i).
\end{align*}
Here, $(m_{i\to a}(x))_{(i,a)\in E}$ and $(m_{a\to i}(x))_{(i,a)\in E}$ are any probability measures on $\mathcal{X}$ satisfying
\begin{equation}
\begin{split}
m_{i\to a}(x) &\propto h_i(x)\prod_{a'\in\partial i\setminus\{a\}} m_{a'\to i}(x)\\
m_{a\to i}(x) &\propto \sum_{\bm{x}_{\partial a}\in\mathcal{S}_a, x_i=x} f_a(\bm{x}_{\partial a}) \prod_{j\in\partial a\setminus\{i\}} m_{j\to a}(x_j).
\end{split}
\label{eq:bpfixed}
\end{equation}
\end{lemma}
Although an efficient algorithm finding the exact minimum the Bethe free energy has not been known,
 one can consider the following simple heuristic algorithm
which tries to find the minimum of the Bethe free energy.

\begin{definition}[Belief propagation (BP)]
Belief propagation is a message-passing algorithm starting from some initial condition $(m^{(0)}_{a\to i}(x))_{(i,a)\in E}$
in which messages are updated according to the following rules
for $t=1,2,\dotsc$
\begin{align*}
m_{i\to a}^{(t)}(x) &\propto h_i(x_i)\prod_{a'\in\partial i\setminus\{a\}} m_{a'\to i}^{(t-1)}(x)\\
m_{a\to i}^{(t)}(x) &\propto \sum_{\bm{x}_{\partial a}\in\mathcal{S}_a, x_i=x} f_a(\bm{x}_{\partial a}) \prod_{j\in\partial a\setminus\{i\}} m_{j\to a}^{(t-1)}(x_j).
\end{align*}
\end{definition}
If BP converges to a fixed point, which is not guaranteed, then one obtains a stationary point of the Bethe free energy.
Also, the fixed point is not necessarily the global minimum of the Bethe free energy.

By elementary calculations, the following clear representation of the Bethe free energy is obtained at stationary points.
\begin{lemma}
For $((b_i)_{i\in V}, (b_a)_{a\in F})\in\mathcal{L}(G)$ satisfying the stationary condition, it holds that
\begin{align*}
&\mathcal{F}_{\mathrm{Bethe}}((b_i)_{i\in V}, (b_a)_{a\in F}) =
-\sum_{a\in F}\log Z_a((m_{i\to a})_{i\in\partial a})\\
&-\sum_{i\in V}\log Z_i((m_{a\to i})_{a\in\partial i})
+\sum_{(i,a)\in E}\log Z_{i,a}(m_{i\to a}, m_{a\to i})
\end{align*}
and hence
\begin{align}
&Z_{\mathrm{Bethe}}((b_i)_{i\in V}, (b_a)_{a\in F}) =
\prod_{a\in F} Z_a((m_{i\to a})_{i\in\partial a})\nonumber\\
&\cdot \prod_{i\in V} Z_i((m_{a\to i})_{a\in\partial i})
\prod_{(i,a)\in E} \frac1{Z_{i,a}(m_{i\to a}, m_{a\to i})}.
\label{eq:Zbethep}
\end{align}
The condition~\eqref{eq:bpfixed} is also the stationary condition for the right-hand sides of the above two equations~\cite[Proposition 14.8]{mezard2009ipa}.
\end{lemma}
Let $\mathrm{IS}(\mathcal{F}_{\mathrm{Bethe}})$ be the set of stationary points of the Bethe free energy in the interior
\begin{align*}
&\bigl\{((b_i\in\mathcal{P}(\mathcal{X}))_{i\in V}, (b_a\in\mathcal{P}(\mathcal{S}_a))_{a\in F})\in\mathcal{L}(G)\mid\\
& b_i(x_i)>0, \forall i\in V, x_i\in\mathcal{X},\,
b_a(\bm{x}_{\partial a})>0,\forall a\in F, \bm{x}_{\partial a}\in\mathcal{S}_a\bigr\}
\end{align*}
of $\mathcal{L}(G)$.

Note that there is a dual definition of the Bethe approximation using the Legendre transformation from the log-partition function to the minus entropy
while the above definition uses the Legendre transformation from the minus entropy to the log-partition function~\cite{ikeda2004stochastic}, \cite{watanabe2010thesis}, \cite{Werner-UAI-2010}.
Although it is also an interesting characterization of the Bethe approximation,
we only introduce the following result, which is related to the dual definition, and is easily confirmed 
by using the equations in Lemma~\ref{lem:bstat}.

\begin{lemma}[\cite{wainwright2003tree}]\label{lem:repara}
For any $((b_i)_{i\in V}, (b_a)_{a\in F})\in\mathrm{IS}(\mathcal{F}_{\mathrm{Bethe}})$, it holds
\begin{align}
&\prod_{a\in F}f_a(\bm{x}_{\partial a})\prod_{i\in V} h_i(x_i) = Z_{\mathrm{Bethe}}((b_i)_{i\in V}, (b_a)_{a\in F})\nonumber\\
&\qquad\cdot\prod_{a\in F}\frac{b_a(\bm{x}_{\partial a})}{\prod_{i\in\partial a}b_i(x_i)}\prod_{i\in V} b_i(x_i),
\hspace{2em} \text{for } \bm{x}\in\mathcal{X}^N
\nonumber\\
&Z(G) = Z_{\mathrm{Bethe}}((b_i)_{i\in V}, (b_a)_{a\in F})\nonumber\\
&\qquad\cdot\sum_{\bm{x}\in\mathcal{X}^N}\prod_{a\in F}\frac{b_a(\bm{x}_{\partial a})}{\prod_{i\in\partial a}b_i(x_i)}\prod_{i\in V} b_i(x_i).
\label{eq:repara}
\end{align}
\end{lemma}
The equation~\eqref{eq:LC} can be proved by expanding the right-hand side of~\eqref{eq:repara}~\cite{sudderth2008loop}.
On the other hand, the proof of~\eqref{eq:LC} based on a general equality was shown in~\cite{chernyak2007loop},
in which the Bethe approximations at stationary points naturally appear on some conditions.
In other word, the proof of~\eqref{eq:LC} in~\cite{chernyak2007loop} gives a new characterization of the Bethe approximation.
In the next section,~\eqref{eq:LC} is generalized using the idea shown in~\cite{chernyak2007loop}.

\section{Holographic transformation and loop calculus for finite alphabet}\label{sec:hol}
\subsection{Holographic transformation and Holant theorem}\label{subsec:hol}
\begin{figure}
\begin{center}
\subfloat[]{
\begin{tikzpicture}
[scale=0.3, inner sep=0mm, C/.style={circle,minimum size=3mm,draw=black,thick},
S/.style={rectangle,minimum size=3mm,draw=black,thick}, label distance=1mm]
\node (i1) at (0,0) [C,label=above:$h_i$] {};
\node (a1) at (4,0) [S,label=above:$f_a$] {};
\draw (i1) to (a1) [color=black,thick];
\end{tikzpicture}
\label{subfig:a}
}
\hspace{4em}
\subfloat[]{
\begin{tikzpicture}
[scale=0.3, inner sep=0mm, C/.style={circle,minimum size=3mm,draw=black,thick},
S/.style={rectangle,minimum size=3mm,draw=black,thick}, label distance=1mm]
\node (i1) at (0,0) [C,label=above:$h_i$] {};
\node (a1) at (8,0) [S,label=above:$f_a$] {};
\node (i2) at (6,0) [C] {};
\node (a2) at (4,0) [S,label=above:$\delta$] {};
\draw (i1) to (a2) [color=black,thick];
\draw (a2) to (i2) [color=black,thick];
\draw (i2) to (a1) [color=black,thick];
\end{tikzpicture}
\label{subfig:b}
}
\hspace{4em}
\subfloat[]{
\begin{tikzpicture}
[scale=0.3, inner sep=0mm, C/.style={circle,minimum size=3mm,draw=black,thick},
S/.style={rectangle,minimum size=3mm,draw=black,thick}, label distance=1mm]
\node (i1) at (0,0) [C,very thick, fill=black,label=above:$h_i$] {};
\node (a1) at (10,0) [S,label=above:$f_a$] {};
\node (i2) at (8,0) [C,very thick, fill=black] {};
\node (a2) at (2,0) [S,label=above:$\phi_{i,a}$] {};
\node (i3) at (4,0) [C] {};
\node (a3) at (6,0) [S,label=above:$\hat{\phi}_{i,a}$] {};
\draw (i1) to (a2) [color=black,thick];
\draw (a2) to (i3) [color=black,thick];
\draw (i3) to (a3) [color=black,thick];
\draw (a3) to (i2) [color=black,thick];
\draw (i2) to (a1) [color=black,thick];
\end{tikzpicture}
\label{subfig:c}
}
\hspace{4em}
\subfloat[]{
\begin{tikzpicture}
[scale=0.3, inner sep=0mm, C/.style={circle,minimum size=3mm,draw=black,thick},
S/.style={rectangle,minimum size=3mm,draw=black,thick}, label distance=1mm]
\node (i1) at (0,0) [S,label=above:$\hat{h}_i$] {};
\node (a1) at (8,0) [S,label=above:$\hat{f}_a$] {};
\node (i2) at (4,0) [C] {};
\draw (i1) to (i2) [color=black,thick];
\draw (i2) to (a1) [color=black,thick];
\end{tikzpicture}
\label{subfig:d}
}
\end{center}
\caption{A graphical explanation of the Holant theorem on the condition $|\mathcal{Y}|\ge|\mathcal{X}|$.
 (a) A pair of connected variable node and factor node in a factor graph.
(b) The equality constraint and new variable node are inserted to an edge.
(c) The equality constraint is separated into $\phi_{i,a}$ and $\hat{\phi}_{i,a}$.
Here, for every edge in the factor graph, an original edge (a) is transformed to (c).
Then, the summations for all of the filled variable nodes in the factor graph are taken.
(d) The new representation for the partition function is obtained.
}
\label{fig:holant}
\end{figure}
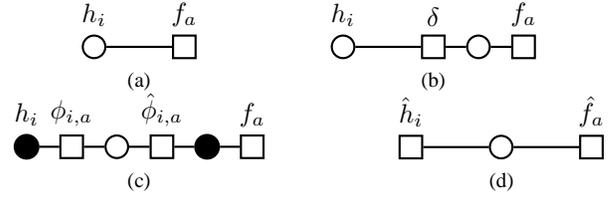

To introduce loop calculus, we use the idea of local linear transformations in~\cite{chernyak2007loop}.
This idea can be recognized as a holographic transformation (also called gauge transformation in physics).
Holographic transformations were introduced by Valiant~\cite{valiant2008holographic} and simplified in~\cite{5695119}. 
The explanation of~\eqref{eq:LC} by holographic transformation is also mentioned in~\cite{forney2011partition}.
First, we assume that $f_a(\bm{x}_{\partial a})$ has the following representation
\begin{equation}
f_a(\bm{x}_{\partial a}) = \sum_{\bm{y}_{\partial a}\in\mathcal{Y}^{d_a}} \hat{f}_a(\bm{y}_{\partial a}) \prod_{i\in\partial a} \phi_{i,a}(x_i, y_i)
\label{eq:hol0}
\end{equation}
for each $a\in F$ for some set $\mathcal{Y}$ and $\phi_{i,a}\colon \mathcal{X}\times\mathcal{Y}\to \mathbb{R}$.
Generally, this representation can be obtained when $|\mathcal{Y}|\ge|\mathcal{X}|$ by letting
\begin{equation}
\hat{f}_a(\bm{y}_{\partial a}) = \sum_{\bm{x}_{\partial a}\in\mathcal{X}^{d_a}} f_a(\bm{x}_{\partial a}) \prod_{i\in\partial a} \hat{\phi}_{i,a}(y_i, x_i)
\label{eq:hol1}
\end{equation}
for some $\hat{\phi}_{i,a}\colon\mathcal{Y}\times\mathcal{X}\to\mathbb{R}$
where $\phi_{i,a}$ and $\hat{\phi}_{i,a}$ satisfy
\begin{equation}
\sum_{y\in\mathcal{Y}} \phi_{i,a}(x, y)\hat{\phi}_{i,a}(y, z) = \delta(x,z)
\label{eq:inv0}
\end{equation}
for all $(i,a)\in E$.
The linear transformation~\eqref{eq:hol1} is called the holographic transformation.
When $|\mathcal{Y}|<|\mathcal{X}|$, there is no choice of $\phi_{i,a}$ and $\hat{\phi}_{i,a}$ satisfying~\eqref{eq:inv0}.
Even for this latter case, if it holds that
\begin{equation*}
f_a(\bm{x}_{\partial a}) = \sum_{\bm{z}_{\partial a}\in\mathcal{X}^{d_a}} f_a(\bm{z}_{\partial a}) \prod_{i\in\partial a} \psi_{i,a}(x_i, z_i)
\end{equation*}
where $\psi_{i,a}(x,z):= \sum_{y\in\mathcal{Y}} \phi_{i,a}(x, y)\hat{\phi}_{i,a}(y, z)$,
i.e., $f_a(\bm{x}_{\partial a})$ is an eigenvector of $\prod_{i\in\partial a}\psi_{i,a}(x,z)$ corresponding to an eigenvalue 1, it also holds~\eqref{eq:hol0} and \eqref{eq:hol1}.
When we have the representation~\eqref{eq:hol0} of $f_a(\bm{x}_{\partial a})$ for all $a\in F$, one obtains
\begin{align*}
&Z(G) = \sum_{\bm{x}\in\mathcal{X}^N}  \prod_{a\in F}f_a(\bm{x}_{\partial a})\prod_{i\in V} h_i(x_i)\\
&= \sum_{\bm{x}\in\mathcal{X}^N} 
 \prod_{a\in F}\left(\sum_{\bm{y}_{\partial a}\in\mathcal{Y}^{d_a}} \hat{f}_a(\bm{y}_{\partial a}) \prod_{i\in\partial a} \phi_{i,a}(x_i, y_i)\right)
\prod_{i\in V} h_i(x_i)\\
&= \sum_{\bm{x}\in\mathcal{X}^N} \sum_{\bm{y}\in\mathcal{Y}^{|E|}}
 \prod_{a\in F}\left(\hat{f}_a(\bm{y}_{\partial a, a}) \prod_{i\in\partial a} \phi_{i,a}(x_i, y_{i,a})\right)
\prod_{i\in V} h_i(x_i)\\
&= \sum_{\bm{y}\in\mathcal{Y}^{|E|}} \prod_{a\in F} \hat{f}_a(\bm{y}_{\partial a, a})
\prod_{i\in V}
\left(\sum_{x\in\mathcal{X}} h_i(x)\prod_{a\in\partial i}\phi_{i,a}(x,y_{i,a})\right).
\end{align*}
Here, $\bm{y}_{\partial a, a}:= (y_{i,a})_{i\in\partial a}$.
By letting
\begin{align*}
\hat{h}_i(\bm{y}_{i,\partial i}) := \sum_{x\in\mathcal{X}} h_i(x)\prod_{a\in\partial i}\phi_{i,a}(x,y_{i,a})
\end{align*}
one obtains
\begin{equation}
Z(G) = \sum_{\bm{y}\in\mathcal{Y}^{|E|}} \prod_{a\in F} \hat{f}_a(\bm{y}_{\partial a, a})
\prod_{i\in V} \hat{h}_i(\bm{y}_{i,\partial i})
\label{eq:holant}
\end{equation}
where $\bm{y}_{i, \partial i}:= (y_{i,a})_{a\in\partial i}$.
The equation~\eqref{eq:holant} is called the Holant theorem in~\cite{valiant2008holographic}, \cite{5695119}.
A graphical explanation of the Holant theorem is shown in Fig.~\ref{fig:holant}.
Even if the original weights $f_a$ and $h_i$ are non-negative, new weights $\hat{f}_a$ and $\hat{h}_i$ are not necessarily non-negative.
Note that~\eqref{eq:holant} holds on any commutative ring once one has~\eqref{eq:hol0}.
Many equalities including~\eqref{eq:LC}
in information theory, machine learning and computer science can be understood by the Holant theorem~\cite{forney2011partition}.

\subsection{Loop calculus for the binary alphabet}\label{subsec:lc}
In the following, we assume $\mathcal{Y}=\mathcal{X}$.
In this case, the condition~\eqref{eq:inv0} is equivalent to
\begin{equation}
\sum_{x\in\mathcal{X}} \hat{\phi}_{i,a}(y, x)\phi_{i,a}(x, w) = \delta(y,w)
\label{eq:inv1}
\end{equation}
i.e., $\phi_{i,a}$ represents the inverse matrix of a matrix represented by $\hat{\phi}_{i,a}$.
Let $\underline{0}$ be an arbitrary fixed element in $\mathcal{X}$.
In~\eqref{eq:holant}, variables assigned not to $\underline{0}$ is regarded as chosen edges.
In order to fix $\phi_{i,a}$ and $\hat{\phi}_{i,a}$ explicitly for the loop calculus, we employ the following additional conditions
for all $i\in V$ and $a\in F$
\begin{align}
\hat{f}_a(\bm{y}_{\partial a,a}) = 0,&\quad\text{when } y_{i,a} = \underline{0} \text{ for all but one } i\in\partial a\label{eq:ac0}\\
\hat{h}_i(\bm{y}_{i,\partial i}) = 0,&\quad\text{when } y_{i,a} = \underline{0} \text{ for all but one } a\in\partial i\label{eq:ac1}.
\end{align}
The conditions~\eqref{eq:ac0} and \eqref{eq:ac1} are considered in~\cite{chernyak2007loop}.
When the conditions~\eqref{eq:ac0} and \eqref{eq:ac1} are satisfied, only $\bm{y}\in\mathcal{X}^{|E|}$ satisfying
\begin{align*}
&\{(i,a)\in E\mid y_{i,a}\ne\underline{0} \}\\
&\in
\{E'\subseteq E\mid d_i(E')\ne 1,\, \forall i\in V, d_a(E')\ne 1,\, \forall a \in F\}=:\mathcal{G}
\end{align*}
 can have non-zero weight in~\eqref{eq:holant}
where $d_i(E'):=|\{a\in F\mid (i,a)\in E'\}|$ and $d_a(E'):=|\{i\in V\mid (i,a)\in E'\}|$.
Elements in $\mathcal{G}$ are called generalized loops.
\begin{figure}
\begin{center}
\begin{tikzpicture}
[scale=0.3, inner sep=1mm, C/.style={circle,draw=black,thick},
S/.style={rectangle,draw=black,thick}]
\node (a3) at (0,0) [S] {};
\node (a2) at (4,0) [S] {};
\node (1) at (0,{2/cos(30)}) [C] {};
\node (2) at (4,{2/cos(30)}) [C] {};
\node (3) at (2,{2/cos(30)-4*sin(60)}) [C] {};
\draw (a3) to (3) [color=black,thick];
\draw (a3) to (1) [color=black,thick];
\draw (a2) to (2) [color=black,thick];
\draw (a2) to (3) [color=black,thick];
\node (z2) at (6,{4*sin(60)}) [S] {};
\node (z22) at (8.5,{4*sin(60)}) [C] {};
\node (z222) at (8.5,{1.5*sin(60)}) [C] {};
\draw (2) to (z2) [color=black,thick];
\draw (z2) to (z22) [color=black,thick];
\draw (z2) to (z222) [color=black,thick];
\node (z3) at (-2.5,1.5) [C] {};
\node (z33) at (-5,2.5) [S] {};
\node (z4) at (-2.5,{4*sin(60)}) [S] {};
\draw (z4) to (z3) [color=black,thick];
\draw (z4) to (1) [color=black,thick];
\draw (a3) to (z3) [color=black,thick];
\draw (z3) to (z33) [color=black,thick];
\draw (z222) to (a2) [color=black,thick];
\node (4) at (-5,0.5) [C] {};
\draw (4) to (z33) [color=black,thick];
\end{tikzpicture}
\\
\begin{tikzpicture}
[scale=0.3, inner sep=1mm, C/.style={circle,draw=black,thick},
S/.style={rectangle,draw=black,thick}]
\node (a3) at (0,0) [S] {};
\node (a2) at (4,0) [S] {};
\node (1) at (0,{2/cos(30)}) [C] {};
\node (2) at (4,{2/cos(30)}) [C] {};
\node (3) at (2,{2/cos(30)-4*sin(60)}) [C] {};
\draw (a3) to (3) [color=black,thick];
\draw (a3) to (1) [color=red,very thick];
\draw (a2) to (2) [color=black,thick];
\draw (a2) to (3) [color=black,thick];
\node (z2) at (6,{4*sin(60)}) [S] {};
\node (z22) at (8.5,{4*sin(60)}) [C] {};
\node (z222) at (8.5,{1.5*sin(60)}) [C] {};
\draw (2) to (z2) [color=black,thick];
\draw (z2) to (z22) [color=black,thick];
\draw (z2) to (z222) [color=black,thick];
\node (z3) at (-2.5,1.5) [C] {};
\node (z33) at (-5,2.5) [S] {};
\node (z4) at (-2.5,{4*sin(60)}) [S] {};
\draw (z4) to (z3) [color=red,very thick];
\draw (z4) to (1) [color=red,very thick];
\draw (a3) to (z3) [color=red,very thick];
\draw (z3) to (z33) [color=black,thick];
\draw (z222) to (a2) [color=black,thick];
\node (4) at (-5,0.5) [C] {};
\draw (4) to (z33) [color=black,thick];
\end{tikzpicture}
\hspace{2em}
\begin{tikzpicture}
[scale=0.3, inner sep=1mm, C/.style={circle,draw=black,thick},
S/.style={rectangle,draw=black,thick}]
\node (a3) at (0,0) [S] {};
\node (a2) at (4,0) [S] {};
\node (1) at (0,{2/cos(30)}) [C] {};
\node (2) at (4,{2/cos(30)}) [C] {};
\node (3) at (2,{2/cos(30)-4*sin(60)}) [C] {};
\draw (a3) to (3) [color=black,thick];
\draw (a3) to (1) [color=black,thick];
\draw (a2) to (2) [color=red,very thick];
\draw (a2) to (3) [color=black,thick];
\node (z2) at (6,{4*sin(60)}) [S] {};
\node (z22) at (8.5,{4*sin(60)}) [C] {};
\node (z222) at (8.5,{1.5*sin(60)}) [C] {};
\draw (2) to (z2) [color=red,very thick];
\draw (z2) to (z22) [color=black,thick];
\draw (z2) to (z222) [color=red,very thick];
\node (z3) at (-2.5,1.5) [C] {};
\node (z33) at (-5,2.5) [S] {};
\node (z4) at (-2.5,{4*sin(60)}) [S] {};
\draw (z4) to (z3) [color=black,thick];
\draw (z4) to (1) [color=black,thick];
\draw (a3) to (z3) [color=black,thick];
\draw (z3) to (z33) [color=black,thick];
\draw (z222) to (a2) [color=red,very thick];
\node (4) at (-5,0.5) [C] {};
\draw (4) to (z33) [color=black,thick];
\end{tikzpicture}
\\
\begin{tikzpicture}
[scale=0.3, inner sep=1mm, C/.style={circle,draw=black,thick},
S/.style={rectangle,draw=black,thick}]
\node (a3) at (0,0) [S] {};
\node (a2) at (4,0) [S] {};
\node (1) at (0,{2/cos(30)}) [C] {};
\node (2) at (4,{2/cos(30)}) [C] {};
\node (3) at (2,{2/cos(30)-4*sin(60)}) [C] {};
\draw (a3) to (3) [color=black,thick];
\draw (a3) to (1) [color=red,very thick];
\draw (a2) to (2) [color=red,very thick];
\draw (a2) to (3) [color=black,thick];
\node (z2) at (6,{4*sin(60)}) [S] {};
\node (z22) at (8.5,{4*sin(60)}) [C] {};
\node (z222) at (8.5,{1.5*sin(60)}) [C] {};
\draw (2) to (z2) [color=red,very thick];
\draw (z2) to (z22) [color=black,thick];
\draw (z2) to (z222) [color=red,very thick];
\node (z3) at (-2.5,1.5) [C] {};
\node (z33) at (-5,2.5) [S] {};
\node (z4) at (-2.5,{4*sin(60)}) [S] {};
\draw (z4) to (z3) [color=red,very thick];
\draw (z4) to (1) [color=red,very thick];
\draw (a3) to (z3) [color=red,very thick];
\draw (z3) to (z33) [color=black,thick];
\draw (z222) to (a2) [color=red,very thick];
\node (4) at (-5,0.5) [C] {};
\draw (4) to (z33) [color=black,thick];
\end{tikzpicture}
\hspace{2em}
\begin{tikzpicture}
[scale=0.3, inner sep=1mm, C/.style={circle,draw=black,thick},
S/.style={rectangle,draw=black,thick}]
\node (a3) at (0,0) [S] {};
\node (a2) at (4,0) [S] {};
\node (1) at (0,{2/cos(30)}) [C] {};
\node (2) at (4,{2/cos(30)}) [C] {};
\node (3) at (2,{2/cos(30)-4*sin(60)}) [C] {};
\draw (a3) to (3) [color=red,very thick];
\draw (a3) to (1) [color=red,very thick];
\draw (a2) to (2) [color=red,very thick];
\draw (a2) to (3) [color=red,very thick];
\node (z2) at (6,{4*sin(60)}) [S] {};
\node (z22) at (8.5,{4*sin(60)}) [C] {};
\node (z222) at (8.5,{1.5*sin(60)}) [C] {};
\draw (2) to (z2) [color=red,very thick];
\draw (z2) to (z22) [color=black,thick];
\draw (z2) to (z222) [color=red,very thick];
\node (z3) at (-2.5,1.5) [C] {};
\node (z33) at (-5,2.5) [S] {};
\node (z4) at (-2.5,{4*sin(60)}) [S] {};
\draw (z4) to (z3) [color=red,very thick];
\draw (z4) to (1) [color=red,very thick];
\draw (a3) to (z3) [color=red,very thick];
\draw (z3) to (z33) [color=black,thick];
\draw (z222) to (a2) [color=red,very thick];
\node (4) at (-5,0.5) [C] {};
\draw (4) to (z33) [color=black,thick];
\end{tikzpicture}
\end{center}
\caption{Generalized loops on a connected factor graph. 
Five generalized loops are shown.
The sets of red thick edges correspond to generalized loops.}
\label{fig:gloops}
\end{figure}
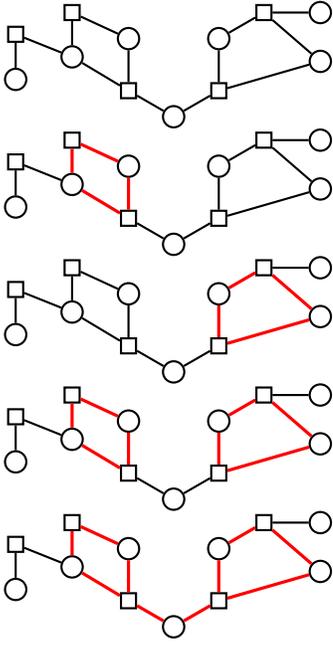
An example of generalized loop is shown in Fig.~\ref{fig:gloops}.
From~\eqref{eq:ac0}, it holds
\begin{equation}
\sum_{\bm{x}_{\partial a}\in\mathcal{X}^{d_a}} \left(f_a(\bm{x}_{\partial a}) \prod_{j\in\partial a\setminus \{i\}} \hat{\phi}_{j,a}(\underline{0}, x_j)\right) \hat{\phi}_{i,a}(y, x_i) = 0
\label{eq:orth0}
\end{equation}
for any $y\in\mathcal{X}\setminus\{\underline{0}\}$ and any $(i,a)\in E$.
The equation~\eqref{eq:orth0} means
\begin{equation}
\sum_{\bm{x}_{\partial a\setminus\{i\}}\in\mathcal{X}^{d_a}} f_a(\bm{x}_{\partial a}) \prod_{j\in\partial a\setminus \{i\}} \hat{\phi}_{j,a}(\underline{0}, x_j)
\label{eq:gbp00}
\end{equation}
must be orthogonal with $[\hat{\phi}_{i,a}(y, x_i)]_{x_i\in\mathcal{X}}$ for all $y\in\mathcal{X}\setminus\{\underline{0}\}$.
Since $q-1$ $q$-dimensional vectors $\{[\hat{\phi}_{i,a}(y, x_i)]_{x_i\in\mathcal{X}}\}_{y\in\mathcal{X}\setminus\{\underline{0}\}}$ are linearly independent,
the $q$-dimensional vector \eqref{eq:gbp00} is uniquely determined up to a constant factor.
From the condition~\eqref{eq:inv1}, the same condition is also required for $\phi_{i,a}(x_i,\underline{0})$.
Hence,~\eqref{eq:gbp00} must be proportional to $\phi_{i,a}(x_i,\underline{0})$.
From the diagonal constraints of~\eqref{eq:inv1}, one obtains
\begin{equation}
\phi_{i,a}(x_i,\underline{0}) = 
\frac1{\hat{f}_a(\underline{\bm{0}})}\sum_{\bm{x}_{\partial a\setminus\{i\}}\in\mathcal{X}^{d_a}} f_a(\bm{x}_{\partial a}) \prod_{j\in\partial a\setminus \{i\}} \hat{\phi}_{j,a}(\underline{0}, x_j)
\label{eq:gbp0}
\end{equation}
where $\underline{\bm{0}}$ is the all-$\underline{0}$ assignment.
For the same reason, from~\eqref{eq:inv1} and \eqref{eq:ac1}, one obtains
\begin{equation}
\hat{\phi}_{i,a}(\underline{0},x) = 
\frac1{\hat{h}_i(\underline{\bm{0}})}h_i(x) \prod_{b\in\partial i\setminus \{a\}} \phi_{i,b}(x, \underline{0}).
\label{eq:gbp1}
\end{equation}
Conversely, if the conditions~\eqref{eq:inv1}, \eqref{eq:gbp0} and \eqref{eq:gbp1} are satisfied then the conditions~\eqref{eq:ac0} and \eqref{eq:ac1} are also satisfied.
Hence, the set of conditions~\eqref{eq:inv1}, \eqref{eq:gbp0} and \eqref{eq:gbp1} is equivalent to the set of conditions~\eqref{eq:inv1}, \eqref{eq:ac0} and \eqref{eq:ac1}. 
In the following, we consider $\phi_{i,a}$ and $\hat{\phi}_{i,a}$ satisfying the conditions~\eqref{eq:inv1}, \eqref{eq:gbp0} and \eqref{eq:gbp1}.
From the conditions~\eqref{eq:gbp0} and \eqref{eq:gbp1}, $(\hat{\phi}_{i,a}(\underline{0},x), \phi_{i,a}(x,\underline{0}))_{(i,a)\in E}$ must be proportional to $(m_{i\to a}(x), m_{a\to i}(x))_{(i,a)\in E}$ which is a fixed point of BP equations~\eqref{eq:bpfixed}.
Although any complex solution of BP equations~\eqref{eq:bpfixed} is allowed, here, only non-negative real solutions are considered for the simplicity.
Then, one obtains
\begin{align}
\phi_{i,a}(x,\underline{0})&=c_{i,a}m_{a\to i}(x),&
\hat{\phi}_{i,a}(\underline{0},x)&=\hat{c}_{i,a}m_{i\to a}(x)
\label{eq:b0}
\end{align}
where $c_{i,a}$ and $\hat{c}_{i,a}$ are arbitrary constants satisfying $c_{i,a}\hat{c}_{i,a}=1/Z_{i,a}(m_{i\to a}, m_{a\to i})$.
The choice of the constants does not change each weight of $\bm{y}\in\mathcal{X}^{|E|}$ in~\eqref{eq:holant} since the constants appear in~\eqref{eq:holant} as the product $c_{i,a}\hat{c}_{i,a}$.
For the same reason,
the transformation
$(\phi_{i,a}(x,y), \hat{\phi}_{i,a}(y,x))_{x\in\mathcal{X}}\to (c_{i,a}(y)\phi_{i,a}(x,y), 1/c_{i,a}(y)\hat{\phi}_{i,a}(y,x))_{x\in\mathcal{X}}$
does not change each of the weight in~\eqref{eq:holant} for any constant $c_{i,a}(y)$ and any $y\in\mathcal{X}$,
and hence we do not have to distinguish
$(\phi_{i,a}(x,y), \hat{\phi}_{i,a}(y,x))_{x\in\mathcal{X}}$ up to the constant factor for each $y\in\mathcal{X}$.
Since it holds
\begin{align*}
\hat{f}_a(\underline{\bm{0}}) &= Z_a\left((m_{i\to a})_{i\in\partial a}\right)\prod_{i\in\partial a} \hat{c}_{i,a}\\
\hat{h}_i(\underline{\bm{0}}) &= Z_i\left((m_{a\to i})_{a\in\partial i}\right)\prod_{a\in\partial i} c_{i,a}
\end{align*}
and from~\eqref{eq:Zbethep},
the weight of the all-$\underline{0}$ assignment in~\eqref{eq:holant} is $Z_{\mathrm{Bethe}}((b_i)_{i\in V}, (b_a)_{a\in F}))$.
In this context, the stationary condition~\eqref{eq:bpfixed} of the Bethe free energy and the Bethe partition function naturally appear only from the conditions~\eqref{eq:ac0} and \eqref{eq:ac1}.
Hence, this story gives a new characterization of the Bethe approximation.
The significant result was obtained in~\cite{chernyak2007loop}.

On the other hand,
there is no constraint for
$(\phi_{i,a}(x,y), \hat{\phi}_{i,a}(y,x))_{x\in\mathcal{X}, y\in\mathcal{X}\setminus\{\underline{0}\}}$
except for~\eqref{eq:inv1}.
Hence, there still exist degrees of freedom for the choices of
$(\phi_{i,a}(x,y), \hat{\phi}_{i,a}(y,x))_{x\in\mathcal{X}, y\in\mathcal{X}\setminus\{\underline{0}\}}$.
For the binary alphabet, i.e., $\mathcal{X}=\{\underline{0}, \underline{1}\}$, the vectors
$(\phi_{i,a}(x,\underline{1}), \hat{\phi}_{i,a}(\underline{1},x))_{x\in\mathcal{X}}$ are uniquely determined up to a constant factor, e.g.,
\begin{equation}
\begin{split}
\phi_{i,a}(x,\underline{1})&=(-1)^{\bar{x}}c_{i,a}m_{i\to a}(\bar{x})\\
\hat{\phi}_{i,a}(\underline{1},x)&=(-1)^{\bar{x}}\hat{c}_{i,a}m_{a\to i}(\bar{x})
\end{split}
\label{eq:b1}
\end{equation}
where $\bar{\underline{0}}:=\underline{1}$ and $\bar{\underline{1}}:=\underline{0}$.
From~\eqref{eq:b0} and \eqref{eq:b1}, Chertkov and Chernyak's result for the binary alphabet is obtained as follows.

\begin{lemma}[Loop calculus for the binary alphabet~\cite{chertkov2006loop}]\label{lem:bloop}
Assume that the alphabet is binary, i.e., $\mathcal{X}=\{\underline{0},\underline{1}\}$.
For any $((b_i)_{i\in V}, (b_a)_{a\in F})\in\mathrm{IS}(\mathcal{F}_{\mathrm{Bethe}})$,
it holds
\begin{equation}
Z(G) = Z_\mathrm{Bethe}((b_i)_{i\in V},(b_a)_{a\in F})\sum_{E'\subseteq E} \mathcal{K}_G(E')
\label{eq:loop}
\end{equation}
where
\begin{align*}
\mathcal{K}_G(E')&:=
\prod_{a\in F}\left\langle\prod_{i\in \partial a,\, (i,a)\in E'}\frac{X_i-\eta^i}{\sqrt{\langle(X_i-\eta^i)^2\rangle_{b_i}}}\right\rangle_{b_a}\\
&\quad\cdot\prod_{i\in V}\left\langle\left(\frac{X_i-\eta^i}{\sqrt{\langle(X_i-\eta^i)^2\rangle_{b_i}}}\right)^{d_i(E')}\right\rangle_{b_i}.
\end{align*}
In the first factor of the weight,
 $(X_i)_{i\in\partial a}$ is a random variable taking $\bm{x}_{\partial a}\in\{0,1\}^{d_a}$
 with probability $b_a(\underline{\bm{x}_{\partial a}})$.
In the second factor of the weight,
 $X_i$ is a binary random variable taking 0 and 1 with probability $b_i(\underline{0})$ and $b_i(\underline{1})$, respectively.
In the above equation, $\eta^i:=\langle X_i\rangle_{b_i}= b_i(\underline{1})$.
\end{lemma}
\begin{IEEEproof}
It holds
\begin{align*}
Z(G)
&= \sum_{\bm{y}\in\{\underline{0},\underline{1}\}^{|E|}} \prod_{a\in F} \hat{f}_a(\bm{y}_{\partial a, a})
\prod_{i\in V} \hat{h}_i(\bm{y}_{i, \partial i})\\
&= Z_{\mathrm{Bethe}} \sum_{\bm{y}\in\{\underline{0},\underline{1}\}^{|E|}}
\prod_{a\in F} \left\langle\prod_{i\in\partial a} \frac{\hat{\phi}_{i,a}(X_i,y_{i,a})}{\hat{\phi}_{i,a}(X_i,\underline{0})}\right\rangle_{b_a}\\
&\quad\cdot\prod_{i\in V} \left\langle \prod_{a\in\partial i} \frac{\phi_{i,a}(X_i,y_{i,a})}{\phi_{i,a}(X_i,\underline{0})}\right\rangle_{b_i}\\
&= Z_{\mathrm{Bethe}} \sum_{E'\subseteq E}
\prod_{a\in F} \left\langle\prod_{i\in\partial a, (i,a)\in E'} \frac{\hat{\phi}_{i,a}(X_i,\underline{1})}{\hat{\phi}_{i,a}(X_i,\underline{0})}\right\rangle_{b_a}\\
&\quad\cdot\prod_{i\in V} \left\langle \prod_{a\in\partial i, (i,a)\in E'} \frac{\phi_{i,a}(X_i,\underline{1})}{\phi_{i,a}(X_i,\underline{0})}\right\rangle_{b_i}\\
&= Z_{\mathrm{Bethe}} \sum_{E'\subseteq E}\\
&\prod_{a\in F} \left\langle\prod_{i\in\partial a, (i,a)\in E'}
\frac{(-1)^{\bar{X}_i} m_{i\to a}(\bar{X}_i)m_{a\to i}(\bar{X}_i)}{m_{i\to a}(\underline{0})m_{i\to a}(\underline{1})}\right\rangle_{b_a}\\
&\cdot \prod_{i\in V} \left\langle \prod_{a\in\partial i, (i,a)\in E'}
\frac{(-1)^{\bar{X}_i} m_{i\to a}(\bar{X}_i)m_{a\to i}(\bar{X}_i)}{m_{a\to i}(\underline{0})m_{a\to i}(\underline{1})}\right\rangle_{b_i}.
\end{align*}
Finally,~\eqref{eq:pseudo}, $X_i - \eta^i= (-1)^{\bar{X}_i} b_i(\bar{X}_i)$ and $\langle(X_i-\eta^i)^2\rangle_{b_i}=b_i(\underline{0})b_i(\underline{1})$
complete the proof.
\end{IEEEproof}
Obviously, $X_i$ is not necessarily $(0,1)$-binary random variable and can be regarded as any $(u_i,v_i)$-binary random variable for $u_i\ne v_i$
for every $i\in V$.

\subsection{Loop calculus for non-binary finite alphabets}\label{subsec:nblc}
For non-binary alphabets, as mentioned before,
there still exist degrees of freedom for the choices of
$(\phi_{i,a}(x,y),\allowbreak \hat{\phi}_{i,a}(y,x)\allowbreak)_{x\in\mathcal{X},\allowbreak y\in\mathcal{X}\setminus\{\underline{0}\}}$.
When the fixed point of the BP equation chosen for~\eqref{eq:b0} is in $\mathrm{IS}(\mathcal{F}_{\mathrm{Bethe}})$,
without loss of generality, 
$(\phi_{i,a}(x,y),\allowbreak \hat{\phi}_{i,a}(y,x))_{x\in\mathcal{X}, y\in\mathcal{X}\setminus\{\underline{0}\}}$
can be written in the form
\begin{align*}
\phi_{i,a}(x,y)&=\phi_{i,a}(x,\underline{0}) A^{i,a}_y(x)\\
\hat{\phi}_{i,a}(y,x)&=\hat{\phi}_{i,a}(\underline{0},x) B^{i,a}_y(x)
\end{align*}
for some $(A^{i,a}_y(x))_{x\in\mathcal{X}, y\in\mathcal{X}\setminus\{\underline{0}\}}$ and $(B^{i,a}_y(x))_{x\in\mathcal{X}, y\in\mathcal{X}\setminus\{\underline{0}\}}$ since $\phi_{i,a}(x,\underline{0})$ and $\hat{\phi}_{i,a}(\underline{0},x)$ are non-zero for all $(i,a)\in E$ and $x\in\mathcal{X}$.
Then, the condition~\eqref{eq:inv1} is equivalent to a set of conditions
\begin{equation}
\begin{split}
\sum_{x\in\mathcal{X}} \hat{\phi}_{i,a}(\underline{0},x)\phi_{i,a}(x, y)&=
\left\langle A^{i,a}_y(X_i) \right\rangle_{b_i}=0\\
\sum_{x\in\mathcal{X}} \hat{\phi}_{i,a}(y,x)\phi_{i,a}(x, \underline{0})&=
\left\langle B^{i,a}_y(X_i) \right\rangle_{b_i}=0
\end{split}\label{eq:zeroe}
\end{equation}
for $y\in\mathcal{X}\setminus\{\underline{0}\}$ and
\begin{equation}
\sum_{x\in\mathcal{X}} \hat{\phi}_{i,a}(y,x)\phi_{i,a}(x, w)=
\left\langle A^{i,a}_w(X_i) B^{i,a}_y(X_i) \right\rangle_{b_i}
=\delta(y, w)
\label{eq:duale}
\end{equation}
for $y, w\in\mathcal{X}\setminus\{\underline{0}\}$.
From the conditions~\eqref{eq:zeroe}, $A^{i,a}_y(x)$ and $B^{i,a}_y(x)$ can be written in the forms
\begin{align}
A^{i,a}_y(x)&=\frac{\partial \log b_i(x)}{\partial \eta_{y}^{i,a}},&
B^{i,a}_y(x)&=\frac{\partial \log b_i(x)}{\partial \theta_{y}^{i,a}}
\label{eq:AB}
\end{align}
where $b_i$ is regarded as a point of two parametric families with parameters $\bm{\theta}^{i,a}$ and $\bm{\eta}^{i,a}$ representing an open set of $\mathcal{P}(\mathcal{X})$
(See also~\cite[Section 2.5]{amari2000methods}).
Obviously, the conditions~\eqref{eq:zeroe} are satisfied by~\eqref{eq:AB}.
Conversely, any full-rank $A^{i,a}_y(x)$ and $B^{i,a}_y(x)$ satisfying~\eqref{eq:zeroe} can be written in the forms~\eqref{eq:AB}
since both of them have the degrees of freedom represented by invertible $(q-1)\times(q-1)$ real matrix.
On the forms~\eqref{eq:AB}, the condition~\eqref{eq:duale} is
\begin{equation}
\left\langle 
\frac{\partial \log b_i(X_i)}{\partial \theta_{y}^{i,a}}
\frac{\partial \log b_i(X_i)}{\partial \eta_{w}^{i,a}} \right\rangle_{b_i}
=\delta(y, w).
\label{eq:nb2}
\end{equation}
This condition implies that $\bm{\theta}^{i,a}$ and $\bm{\eta}^{i,a}$ are affine coordinate systems for dual connections~\cite[Section 3.3]{amari2000methods}.
Indeed, the natural parameter $\bm{\theta}^{i,a}$ and the expectation parameter $\bm{\eta}^{i,a}$ for an exponential family satisfies~\eqref{eq:nb2}, which is shown in~\eqref{eq:dual} in Appendix~\ref{apx:ef}.
The following theorem is obtained
from~\eqref{eq:b0} and
\begin{equation}
\begin{split}
\phi_{i,a}(x,y)&=\phi_{i,a}(x,\underline{0})\frac{\partial \log b_i(x)}{\partial \eta_{y}^{i,a}}\\
\hat{\phi}_{i,a}(y,x)&=\hat{\phi}_{i,a}(\underline{0},x)\frac{\partial \log b_i(x)}{\partial \theta_{y}^{i,a}}
\end{split}
\label{eq:b2}
\end{equation}
where $\bm{\theta}^{i,a}$ and $\bm{\eta}^{i,a}$ are the natural parameter and the expectation parameter, respectively,
of $b_i$ as a $(q-1)$-dimensional exponential family representing $\mathcal{P}(\mathcal{X})$ using arbitrary chosen sufficient statistic $\bm{t}^{i,a}(x)$.

\begin{theorem}[Loop calculus for non-binary finite alphabets]\label{thm:non-binary}
For any $((b_i)_{i\in V}, (b_a)_{a\in F})\in\mathrm{IS}(\mathcal{F}_{\mathrm{Bethe}})$,~\eqref{eq:loop} holds
where
\begin{align}
\mathcal{K}_G(E')&:=\sum_{\bm{y}_{E'}\in(\mathcal{X}\setminus\{\underline{0}\})^{|E'|}}
\prod_{a\in F}
\left\langle \prod_{i\in\partial a, (i,a)\in E'} \frac{\partial \log b_i(X_i)}{\partial \theta_{y_{i,a}}^{i,a}}\right\rangle_{b_a}\nonumber\\
&\quad\cdot\prod_{i\in V}
\left\langle \prod_{a\in\partial i, (i,a)\in E'} \frac{\partial \log b_i(X_i)}{\partial \eta_{y_{i,a}}^{i,a}}\right\rangle_{b_i}.
\label{eq:nbloop}
\end{align}
\end{theorem}
\begin{IEEEproof}
Similarly to the proof of Lemma~\ref{lem:bloop}, one obtains
\begin{align*}
Z
&= Z_{\mathrm{Bethe}} \sum_{\bm{y}\in\mathcal{X}^{|E|}}
\prod_{a\in F} \left\langle\prod_{i\in\partial a} \frac{\hat{\phi}_{i,a}(X_i,y_{i,a})}{\hat{\phi}_{i,a}(X_i,\underline{0})}\right\rangle_{b_a}\\
&\quad\cdot\prod_{i\in V} \left\langle \prod_{a\in\partial i} \frac{\phi_{i,a}(X_i,y_{i,a})}{\phi_{i,a}(X_i,\underline{0})}\right\rangle_{b_i}.
\end{align*}
The equation~\eqref{eq:nbloop} is obtained by substituting~\eqref{eq:b2} into the above formula.
\end{IEEEproof}
\begin{remark}
Although this paper only uses algebraic aspects of the tangent vectors of exponential family of distributions on a finite set,
since the relationship~\eqref{eq:nb2} plays a key role in the theory of duality in information geometry,
and since the representation using the tangent vectors for loop calculus 
gives meaning of weights as in the next lemma and also as in Section~\ref{sec:simple},
the author would like to emphasize that the concepts from information geometry plays an important role in this paper.
\end{remark}

Let us confirm that
for any choice of
$(\phi_{i,a}(x,y),\allowbreak \hat{\phi}_{i,a}(y,x))_{x\in\mathcal{X}, y\in\mathcal{X}\setminus\{\underline{0}\}}$
satisfying~\eqref{eq:inv1}, there exists the choice of sufficient statistics such that
\eqref{eq:b2} holds, i.e.,~\eqref{eq:b2} covers all linear transformations satisfying the conditions~\eqref{eq:inv1}, \eqref{eq:ac0} and \eqref{eq:ac1}.
Let $L^{i,a}$ be an arbitrary invertible $(q-1)\times(q-1)$ real (or possibly complex as mentioned at the end of this subsection) matrix whose rows and columns are indexed by $\mathcal{X}\setminus\{\underline{0}\}$, and let
\begin{equation}\label{eq:trans}
t_{y}^{'i,a}(x) := \sum_{w\in\mathcal{X}\setminus\{\underline{0}\}} L^{i,a}_{y,w} t_{w}^{i,a}(x)
\end{equation}
for all $y\in\mathcal{X}\setminus\{\underline{0}\}$.
Since any function $t(x)\colon \mathcal{X}\to\mathbb{R}$ can be represented by linear combination of $\{e_z(x):=\mathbb{I}\{x=z\}\mid z\in\mathcal{X}\setminus\{\underline{0}\}\}$
up to translations, we only have to consider linear transformations between sufficient statistics.
The linear transformation of sufficient statistic affects to the coordinate systems as
\begin{align*}
\theta_{y}^{'i,a} &= \sum_{w\in\mathcal{X}\setminus\{\underline{0}\}} R^{i,a}_{y,w} \theta_{w}^{i,a},&
\eta_{y}^{'i,a} &= \sum_{w\in\mathcal{X}\setminus\{\underline{0}\}} L^{i,a}_{y,w} \eta_{w}^{i,a}
\end{align*}
where $R^{i,a}:= (L^{i,a \, -1})^t$.
The linear transformation of sufficient statistic affects to the tangent vectors as
\begin{align}
\frac{\partial \log b_i(x)}{\partial \eta^{'i,a}_{y}} &= \sum_{w\in\mathcal{X}\setminus\{\underline{0}\}} \frac{\partial \log b_i(x)}{\partial \eta^{i,a}_{w}} \frac{\partial \eta^{i,a}_{w}}{\partial \eta^{'i,a}_{y}}\nonumber\\
&= \sum_{w\in\mathcal{X}\setminus\{\underline{0}\}} \frac{\partial \log b_i(x)}{\partial \eta^{i,a}_{w}} R^{i,a}_{y,w}
\label{eq:t0}
\\
\frac{\partial \log b_i(x)}{\partial \theta^{'i,a}_{y}} &= \sum_{w\in\mathcal{X}\setminus\{\underline{0}\}} \frac{\partial \log b_i(x)}{\partial \theta^{i,a}_{w}} \frac{\partial \theta^{i,a}_{w}}{\partial \theta^{'i,a}_{i,y}}\nonumber\\
&= \sum_{w\in\mathcal{X}\setminus\{\underline{0}\}} \frac{\partial \log b_i(x)}{\partial \theta^{i,a}_{w}} L^{i,a}_{y,w}.
\label{eq:t1}
\end{align}
Hence, the degrees of freedom for
$(\phi_{i,a}(x,y),\allowbreak \hat{\phi}_{i,a}(y,x))_{x\in\mathcal{X}, y\in\mathcal{X}\setminus\{\underline{0}\}}$
in the form of~\eqref{eq:b2} can be represented by $(q-1)\times(q-1)$ invertible matrices.
The degrees of freedom for
$(\phi_{i,a}(x,y),\allowbreak \hat{\phi}_{i,a}(y,x))_{x\in\mathcal{X}, y\in\mathcal{X}\setminus\{\underline{0}\}}$
satisfying~\eqref{eq:b0} and \eqref{eq:inv1} also can be represented by $(q-1)\times(q-1)$ invertible matrices.
Hence, the forms~\eqref{eq:b0} and ~\eqref{eq:b2} can express all choices satisfying the conditions~\eqref{eq:inv1}, \eqref{eq:ac0} and \eqref{eq:ac1}.

From~\eqref{eq:t0} and \eqref{eq:t1}, it can be confirmed that the weight $\mathcal{K}_G(E')$ in~\eqref{eq:nbloop} does not depend on the choice of sufficient statistics as follows
\begin{align*}
&\sum_{\bm{y}_{E'}\in(\mathcal{X}\setminus\{\underline{0}\})^{|E'|}}
\prod_{a\in F}
\left\langle \prod_{i\in\partial a, (i,a)\in E'} \frac{\partial \log b_i(X_i)}{\partial \theta^{'i,a}_{y_{i,a}}}\right\rangle_{b_a}\\
&\quad\cdot\prod_{i\in V}
\left\langle \prod_{a\in\partial i, (i,a)\in E'} \frac{\partial \log b_i(X_i)}{\partial \eta^{'i,a}_{y_{i,a}}}\right\rangle_{b_i}\\
=&\sum_{\bm{y}_{E'}\in(\mathcal{X}\setminus\{\underline{0}\})^{|E'|}}\\
&\prod_{a\in F}
\left\langle \prod_{i\in\partial a, (i,a)\in E'} \left(\sum_{w\in\mathcal{X}\setminus\{\underline{0}\}}\frac{\partial \log b_i(X_i)}{\partial \theta^{i,a}_{w}}L^{i,a}_{y_{i,a}, w}\right)\right\rangle_{b_a}\\
&\cdot\prod_{i\in V}
\left\langle \prod_{a\in\partial i, (i,a)\in E'} \left(\sum_{v\in\mathcal{X}\setminus\{\underline{0}\}}\frac{\partial \log b_i(X_i)}{\partial \eta^{i,a}_{v}}R^{i,a}_{y_{i,a}, v}\right)\right\rangle_{b_i}\\
=&\sum_{\bm{w}_{E'}\in(\mathcal{X}\setminus\{\underline{0}\})^{|E'|}, \bm{v}_{E'}\in(\mathcal{X}\setminus\{\underline{0}\})^{|E'|}}\\
&\prod_{a\in F}
\left\langle \prod_{i\in\partial a, (i,a)\in E'} \frac{\partial \log b_i(X_i)}{\partial \theta^{i,a}_{w_{i,a}}}\right\rangle_{b_a}\\
&\cdot\prod_{i\in V}
\left\langle \prod_{a\in\partial i, (i,a)\in E'} \frac{\partial \log b_i(X_i)}{\partial \eta^{i,a}_{v_{i,a}}}\right\rangle_{b_i}\\
&\cdot \prod_{(i,a)\in E'} \left(\sum_{y\in\mathcal{X}\setminus\{\underline{0}\}} L^{i,a}_{y, w_{i,a}} R^{i,a}_{y, v_{i,a}}\right).
\end{align*}
In the last factor of the above equation,
it holds
$\sum_{y\in\mathcal{X}\setminus\{\underline{0}\}} L^{i,a}_{y, w_{i,a}} R^{i,a}_{y, v_{i,a}}=\delta(w_{i,a}, v_{i,a})$, and hence $\mathcal{K}_G(E')$ is independent of the choice of sufficient statistics.
The above equalities mean that the two representations can be transformed to each other via holographic transformation on $\mathcal{K}_G(E')$.
Note that the weight of generalized loop coincides with what is obtained in~\cite{xiao2011partition} for continuous alphabets.
The details are shown in Appendix~\ref{apx:xiao}.
The expression~\eqref{eq:nbloop} can be further simplified by carefully choosing the sufficient statistics.

\begin{lemma}\label{lem:diag}
If one chooses a common sufficient statistic $\bm{t}^i(x)$ for $\bm{t}^{i,a}(x)$ for each $i\in V$ and all $a\in\partial i$
such that the Fisher information matrix is diagonal at $b_i$ for all $i\in V$, then
it holds
\begin{align*}
&\mathcal{K}_G(E') =\sum_{\bm{y}_{E'}\in(\mathcal{X}\setminus\{\underline{0}\})^{|E'|}}\\
&\prod_{a\in F}
\left\langle \prod_{i\in\partial a, (i,a)\in E'}
\frac{t^i_{y_{i,a}}(X_i)-\eta^i_{y_{i,a}}}{\sqrt{\bigl\langle \bigl(t^i_{y_{i,a}}(X_i)-\eta^i_{y_{i,a}}\bigr)^2\bigr\rangle_{b_i}}}\right\rangle_{b_a}\\
&\quad \cdot
\prod_{i\in V}
\left\langle \prod_{a\in\partial i, (i,a)\in E'}
\frac{t^i_{y_{i,a}}(X_i)-\eta^i_{y_{i,a}}}{\sqrt{\bigl\langle \bigl(t^i_{y_{i,a}}(X_i)-\eta^i_{y_{i,a}}\bigr)^2\bigr\rangle_{b_i}}}\right\rangle_{b_i}.
\end{align*}
\end{lemma}
\begin{IEEEproof}
Let $\bm{\theta}^i$ and $\bm{\eta}^i$ be the natural parameters and the expectation parameters with respect to $\bm{t}^i(x)$.
For the weight of factor nodes, it always holds
\begin{equation*}
\frac{\partial \log b_i(x_i)}{\partial \theta^i_{y_{i,a}}}
=
t^i_{y_{i,a}}(x_i) - \eta^i_{y_{i,a}}.
\end{equation*}
For the weight of variable nodes, it also always holds
\begin{equation*}
\frac{\partial \log b_i(x_i)}{\partial \eta^i_{y_{i,a}}}
=\sum_{w\in\mathcal{X}\setminus \{\underline{0}\}}\frac{\partial \theta^i_{w}}{\partial \eta^i_{y_{i,a}}}
\frac{\partial \log b_i(x_i)}{\partial \theta^i_{w}}.
\end{equation*}
From Appendix~\ref{apx:ef},
$\frac{\partial \theta^i_{w}}{\partial \eta^i_{y_{i,a}}}$ is the $(w,y_{i,a})$-element of the Fisher information matrix $\mathcal{J}_{b_i}(\bm{\eta}^i)$.
When the Fisher information matrix is diagonal, it holds
\begin{align*}
\frac{\partial \log b_i(x_i)}{\partial \eta^i_{y_{i,a}}}
&=
\frac1{\mathcal{J}_{b_i}(\bm{\theta}^i)_{y_{i,a},y_{i,a}}}
\frac{\partial \log b_i(x_i)}{\partial \theta^i_{y_{i,a}}}\\
&=
\frac{t^i_{y_{i,a}}(x_i)-\eta^i_{y_{i,a}}}{\Bigl\langle \bigl(t^i_{y_{i,a}}(X_i)-\eta^i_{y_{i,a}}\bigr)^2\Bigr\rangle_{b_i}}.
\end{align*}
\end{IEEEproof}
Since the Fisher information matrix $\mathcal{J}_{b_i}(\bm{\theta}^i)$ is symmetric real and positive-definite,
there exists an orthogonal matrix $U$ such that $U\mathcal{J}_{b_i}(\bm{\theta}^i)U^{t}=:D$ is a positive diagonal matrix.
By the transformation~\eqref{eq:trans}, the Fisher information matrix is transformed to
$\mathcal{J}_{b_i}({\bm{\theta}'}^i)=L^{i}\mathcal{J}_{b_i}(\bm{\theta}^i)L^{i\,t}$.
Hence, if one chooses $L^{i}=D'VD^{-1/2}U$ for some orthogonal matrix $V$ and some diagonal matrix $D'$,
it holds $\mathcal{J}({\bm{\theta}'}^{i})= D'^2$.
The above discussion shows the existence of the choice of sufficient statistics for the diagonal Fisher information matrix.
The degrees of freedom for the choice of sufficient statistics for Lemma~\ref{lem:diag} are represented by the $(q-1)\times(q-1)$ diagonal matrix $D'$ and the $(q-1)\times(q-1)$ orthogonal matrix  $V$ for each variable node.
Since as mentioned before, a constant multiplication to each $[t^i_{y}(x)]_{x\in\mathcal{X}}$ does not change
the weight of each $\bm{y}_{E'}\in(\mathcal{X}\setminus\{\underline{0}\})^{|E'|}$,
one may regard that the degrees of freedom are represented by a $(q-1)\times(q-1)$ orthogonal matrix for each variable node.

Although the expressions
\begin{equation}
\begin{split}
\phi_{i,a}(x,y)&=\phi_{i,a}(x,\underline{0})\frac{\partial \log b_i(x)}{\partial \theta^{i,a}_{y}}\\
\hat{\phi}_{i,a}(y,x)&=\hat{\phi}_{i,a}(\underline{0},x)\frac{\partial \log b_i(x)}{\partial \eta^{i,a}_{y}}
\end{split}
\label{eq:b3}
\end{equation}
also satisfy~\eqref{eq:inv1},
we use the representation~\eqref{eq:b2} in this paper since~\eqref{eq:b2} gives
a clear representation of weights of certain types of generalized loops as shown in Section~\ref{sec:simple}.
Note that even when one chooses common sufficient statistics for all $i\in V$ and $a\in F$ for~\eqref{eq:b2},
it does not generally hold that the corresponding sufficient statistics for the representation~\eqref{eq:b3} are the common for all $i\in V$ and $a\in F$.

As mentioned in~Section~\ref{subsec:hol},~\eqref{eq:holant} holds on an arbitrary commutative ring.
As a simple example, let us consider the generalization of Theorem~\ref{thm:non-binary} to the complex field.
For given stationary point of the Bethe free energy, complex matrices can be used in~\eqref{eq:trans}.
It gives a simple generalization of Theorem~\ref{thm:non-binary} to the complex field.
More generally, the complex solutions of the BP equation~\eqref{eq:bpfixed} can be used for Theorem~\ref{thm:non-binary}.
Since at least in the author's knowledge, no one has been considered the complex solutions of the BP equation,
it may be an interesting direction of research.
When the local functions $h_i$ and $f_a$ take values in the complex field or finite field,
if one has a solution of the BP equation~\eqref{eq:bpfixed}, one obtains equations similarly to Theorem~\ref{thm:non-binary}
although the degrees of freedom for $(\phi_{i,a}(x,y),\allowbreak \hat{\phi}_{i,a}(y,x))_{x\in\mathcal{X}, y\in\mathcal{X}\setminus\{\underline{0}\}}$ cannot be expressed by using tangent vectors on the information manifold.
In that case, there also exists a difficulty that the constants $Z_a$, $Z_i$ and $Z_{i,a}$ can be zero.
This problem may also appear in the recursive approach of loop calculus for non-binary alphabets~\cite{chernyak2007loop}.
Some techniques used in statistical physics, e.g., Hubbard-Stratonovich transformation and
representation of the equality constraint by an integral of auxiliary variable~\cite{1742-5468-2006-06-P06009},
may be regarded as a holographic transformation using $\mathcal{Y}=\mathbb{R}$.
Relationship between the holographic transformation and the field theory in physics seems interesting.

\subsection{Loop calculus for marginal distributions}\label{subsec:lmarg}
In this subsection, Lemma~\ref{lem:bloop} and Theorem~\ref{thm:non-binary} are generalized to marginal distributions.
Let $g\colon \mathcal{X}^{|C|}\to \mathbb{C}$ be an arbitrary function for $C\subseteq V$.
Then, it holds
\begin{equation*}
Z(G) \langle g(X_C)\rangle_p = \sum_{\bm{x}\in\mathcal{X}^N} g(\bm{x}_C)\prod_{a\in F}f_a(\bm{x}_{\partial a})\prod_{i\in V}h_i(x_i).
\end{equation*}
If $g$ takes a real non-negative value,
$Z(G) \langle g(X_C)\rangle_p$ can be regarded as a partition function of a modified factor graph in which new factor node corresponding to $g$ is added to the original factor graph.
Hence, Lemma~\ref{lem:bloop} and Theorem~\ref{thm:non-binary} can be applied to the modified factor graph.
On the other hand, the same holographic transformation in Lemma~\ref{lem:bloop} and Theorem~\ref{thm:non-binary} can be used for $Z(G) \langle g(X_C)\rangle_p$.
From this idea, the following Lemmas are obtained.
\begin{lemma}\label{lem:bmarginal}
Assume that the alphabet is binary, i.e., $\mathcal{X}=\{\underline{0},\underline{1}\}$.
Let $C\subseteq V$,
$F_C:=\{a\in F\mid \partial a\subseteq C\}$,
$E(F_C):=\{(i,a)\in E\mid a\in F_C\}$
 and $g\colon  \mathcal{X}^{|C|}\to \mathbb{C}$.
For any $((b_i)_{i\in V}, (b_a)_{a\in F})\in\mathrm{IS}(\mathcal{F}_{\mathrm{Bethe}})$,
\begin{align}
&Z(G)\langle g(\bm{X}_{C}) \rangle_p \nonumber\\
&\qquad= Z_\mathrm{Bethe}((b_i)_{i\in V},(b_a)_{a\in F})
\sum_{E'\subseteq E\setminus E(F_C)} \mathcal{K}_G^g(E')
\label{eq:exloop}
\end{align}
where
\begin{align*}
\mathcal{K}_G^g(E')&:=
\prod_{a\in F\setminus F_C}
\left\langle\prod_{i\in \partial a,\, (i,a)\in E'}\frac{X_i-\eta^i}{\sqrt{\langle(X_i-\eta^i)^2\rangle_{b_i}}}\right\rangle_{b_a}\\
&\quad\cdot \prod_{i\in V\setminus C}\left\langle\left(\frac{X_i-\eta^i}{\sqrt{\langle(X_i-\eta^i)^2\rangle_{b_i}}}\right)^{d_i(E')}\right\rangle_{b_i}
\\
&\quad \cdot
\left\langle g(\bm{X}_{C})\prod_{i\in C}\left(\frac{X_i-\eta^i}{\sqrt{\langle(X_i-\eta^i)^2\rangle_{b_i}}}\right)^{d_i(E')}\right\rangle_{b_C}.
\end{align*}
Here, $\langle\cdot\rangle_{b_C}$ is a pseudo-expectation with respect to
an un-normalized distribution
\begin{equation*}
b_C(\bm{x}_{C})
=
\prod_{i\in C} b_i(x_i)\prod_{a\in F_C} \frac{b_a(\bm{x}_{\partial a})}{\prod_{i\in\partial a} b_i(x_i)}.
\end{equation*}
\end{lemma}
\begin{IEEEproof}
By transforming $f_a$ only for $a\notin F_C$, one obtains
\begin{align*}
&Z(G) \langle g(\bm{X}_{C})\rangle_p\\
&= \sum_{\bm{y}_{E\setminus E(F_C)}\in\{0,1\}^{|E\setminus E(F_C)|}} \prod_{a\in F\setminus F_C} \hat{f}_a(\bm{y}_{\partial a})
\prod_{i\in V\setminus C} \hat{h}_i(\bm{y}_{\partial i})\\
&\quad\cdot \Biggl(\sum_{\bm{x}_C} g(\bm{x}_C)\prod_{i\in C} \left(h_i(x_i)\prod_{a\in\partial i, a\notin F_C}\phi_{i,a}(x_i, y_{i,a})\right)\\
&\quad\cdot \prod_{a\in F_C} f_a(\bm{x}_{\partial a})\Biggr).
\end{align*}
One obtains the lemma from the proof of Lemma~\ref{lem:bloop} and
\begin{align*}
&\prod_{i\in C} \left(h_i(x_i) \prod_{a\in\partial i, a\notin F_C} m_{a\to i}(x_i)\right)
\prod_{a\in F_C} f_a(\bm{x}_{\partial a})\\
&=
\prod_{i\in C} Z_i \prod_{a\in F_C} \frac{Z_a}{\prod_{i\in\partial a}Z_{i,a}}
\prod_{i\in C} b_i(x_i)\prod_{a\in F_C} \frac{b_a(\bm{x}_{\partial a})}{\prod_{i\in\partial a} b_i(x_i)}.
\qedhere
\end{align*}
\end{IEEEproof}
The un-normalized distribution $b_C$ is regarded as the Bethe approximation for marginal distribution on $\bm{x}_C$~\cite[Chapter 19]{mezard2009ipa}.
The weight $\mathcal{K}_G^g(\varnothing)$ of the empty graph is the Bethe approximation $\langle g(\bm{X}_C)\rangle_{b_C}$ of expectation of $g(\bm{x}_C)$.
Lemma~\ref{lem:bmarginal} for single variable, i.e., $|C|=1$, was obtained in~\cite{chertkov2006lp} and~\cite{watanabe2010thesis}.
Lemma~\ref{lem:bmarginal} can be generalized to non-binary alphabets as follows.

\begin{lemma}\label{lem:nbmarginal}
Let $C\subseteq V$ and $g\colon  \mathcal{X}^{|C|}\to \mathbb{C}$.
For any $((b_i)_{i\in V}, (b_a)_{a\in F})\in\mathrm{IS}(\mathcal{F}_{\mathrm{Bethe}})$,
it holds~\eqref{eq:exloop}
where
\begin{align*}
\mathcal{K}_G^g(E')&:=\sum_{\bm{y}_{E'}\in(\mathcal{X}\setminus\{\underline{0}\})^{|E'|}}\\
&\quad\prod_{a\in F\setminus F_C}
\left\langle \prod_{i\in\partial a, (i,a)\in E'} \frac{\partial \log b_i(X_i)}{\partial \theta^{i,a}_{y_{i,a}}}\right\rangle_{b_a}\\
&\quad\cdot\left[\prod_{i\in V\setminus C}
\left\langle \prod_{a\in\partial i, (i,a)\in E'} \frac{\partial \log b_i(X_i)}{\partial \eta^{i,a}_{y_{i,a}}}\right\rangle_{b_i}\right]\\
&\quad\cdot\left\langle g(\bm{X}_C)\prod_{i\in C, (i,a)\in E'} \frac{\partial \log b_i(X_i)}{\partial \eta^{i,a}_{y_{i,a}}}\right\rangle_{b_C}.
\end{align*}
If one chooses a common sufficient statistic $\bm{t}^i(x_i)$ for $\bm{t}^{i,a}(x_i)$ for each $i\in V$ and all $a\in\partial i$ such that
the Fisher information matrix is diagonal at $b_i$ for all $i\in V$,
then it holds
\begin{align*}
&\mathcal{K}^g_G(E')=\sum_{\bm{y}_{E'}\in(\mathcal{X}\setminus\{\underline{0}\})^{|E'|}}\\
&\quad\prod_{a\in F\setminus F_C}
\left\langle \prod_{i\in\partial a, (i,a)\in E'}
\frac{t^i_{y_{i,a}}(X_i)-\eta^i_{y_{i,a}}}{\sqrt{\bigl\langle \bigl(t^i_{y_{i,a}}(X_i)-\eta^i_{y_{i,a}}\bigr)^2\bigr\rangle_{b_i}}}\right\rangle_{b_a}
\\
&\quad \cdot
\prod_{i\in V\setminus C}
\left\langle \prod_{a\in\partial i, (i,a)\in E'}
\frac{t^i_{y_{i,a}}(X_i)-\eta^i_{y_{i,a}}}{\sqrt{\bigl\langle \bigl(t^i_{y_{i,a}}(X_i)-\eta^i_{y_{i,a}}\bigr)^2\bigr\rangle_{b_i}}}\right\rangle_{b_i}
\\
&\quad\cdot
\left\langle g(\bm{X}_C)\prod_{i\in C, (i,a)\in E'}
\frac{t^i_{y_{i,a}}(X_i)-\eta^i_{y_{i,a}}}{\sqrt{\bigl\langle \bigl(t^i_{y_{i,a}}(X_i)-\eta^i_{y_{i,a}}\bigr)^2\bigr\rangle_{b_i}}}\right\rangle_{b_C}.
\end{align*}
\end{lemma}
Proof of Lemma~\ref{lem:nbmarginal} is omitted since it is straightforward from the proofs of Theorem~\ref{thm:non-binary} and Lemma~\ref{lem:bmarginal}.
While in Lemma~\ref{lem:bloop} and Theorem~\ref{thm:non-binary}, 
only generalized loops have non-zero weight,
in Lemmas~\ref{lem:bmarginal} and \ref{lem:nbmarginal}, the weight is non-zero only for $E'\in\{E'\subseteq E\setminus E(F_C)\mid d_i(E')\ne 1,\,\forall i\in V\setminus C, d_a(E')\ne 1,\,\forall a\in F\setminus F_C\}$, i.e.,
the weight $\mathcal{K}_G^g(E')$ can be non-zero even if $d_i(E')=1$ for $i\in C$.
Let $\mathbf{1}(\bm{x})=1$ for $\bm{x}\in \mathcal{X}^{|C|}$.
Then, one also obtains from Lemma~\ref{lem:nbmarginal} that
\begin{align}
&\langle g(\bm{X}_C)\rangle_p = \frac{\sum_{E'\subseteq E\setminus E(F_C)} \mathcal{K}_G^g(E')}{\sum_{E'\subseteq E\setminus E(F_C)} \mathcal{K}^{\mathbf{1}}_G(E')}
= \langle g(\bm{X}_C)\rangle_{b_C}\nonumber\\
& + \frac{\sum_{E'\subseteq E\setminus E(F_C)} \left(\mathcal{K}_G^g(E') - \langle g(\bm{X}_C)\rangle_{b_C}\mathcal{K}^{\mathbf{1}}_G(E')\right)}{\sum_{E'\subseteq E\setminus E(F_C)} \mathcal{K}^{\mathbf{1}}_G(E')}.
\label{eq:marg}
\end{align}
This expression is useful for considering a relationship between $\langle g(\bm{X}_C)\rangle_p$ and $\langle g(\bm{X}_C)\rangle_{b_C}$.

\section{Simplifications for simple generalized loops}\label{sec:simple}
Generally, it needs exponential time to take the summation in~\eqref{eq:nbloop}.
In this subsections, simple expressions of weights $\mathcal{K}_G(E')$ and $\mathcal{K}_G^g(E')$ are obtained for some simple $E'\subseteq E$.
They allow efficient computation of the weights.
The weights of a one-dimensional factor graph can be calculated by the transfer matrix method.

\begin{lemma}[Transfer matrix]\label{lem:trans}
Let $S$ be a $q\times q$ matrix whose $(x_1, x_N)$-element is
\begin{equation*}
S_{x_1, x_N}:=
\sum_{(x_2,\dotsc,x_{N-1})\in\mathcal{X}^{N-2}} \prod_{i=1}^{N-1}f_i(x_i, x_{i+1})
\end{equation*}
for $N\ge 2$.
Then, $S=F_1F_2\dotsm F_{N-1}$ where $F_i$ is a $q\times q$ matrix whose $(x, z)$-element is $f_i(x, z)$.
\end{lemma}
\begin{IEEEproof}
The lemma is proved by induction. The lemma is trivial for $N=2$.
The induction step is shown by 
\begin{align*}
S_{x_1, x_N}&=
\sum_{x_2\in\mathcal{X}} f_1(x_1, x_2) 
\sum_{(x_3,\dotsc,x_{N-1})\in\mathcal{X}^{N-3}} \prod_{i=2}^{N-1}f_i(x_i, x_{i+1})\\
&=
\sum_{x_2\in\mathcal{X}} f_1(x_1, x_2) 
\left(F_2F_3\dotsm F_{N-1}\right)_{x_2, x_N}.
\end{align*}
\end{IEEEproof}

As a corollary of Lemma~\ref{lem:trans}, a partition function of a cycle graph in which variables $x_1$ and $x_N$ are identified is $\sum_{x\in\mathcal{X}} S_{x,x} = \tr(F_1\dotsm F_{N-1})$.
Hence, the weight $\mathcal{K}_G(E')$ of a generalized loop $E'\in\{E'\subseteq E\mid d_i(E')\in\{0,2\}, \forall i\in V, d_a(E')\in\{0,2\}\, \forall a\in F, \text{$E'$ is connected}\}=:\mathcal{S}$ can be represented as a trace of the product of matrices.
Here, a generalized loop in $\mathcal{S}$ is called a simple generalized loop.
In the following, we assume that a common sufficient statistic $\bm{t}^i(x)$ is used for $\bm{t}^{i,a}(x)$ in~\eqref{eq:nbloop}.
Then, in~\eqref{eq:nbloop}, the weight of degree-two variable node is an element of the Fisher information matrix $\mathcal{J}_{b_i}(\bm{\eta}^i)$ with respect to the expectation parameter $\bm{\eta}^i$,
which is the inverse matrix of the Fisher information matrix $\mathcal{J}_{b_i}(\bm{\theta}^i)$ with respect to the natural parameter $\bm{\theta}^i$.
Since $(y, w)$-element of the Fisher information matrix $\mathcal{J}_{b_i}(\bm{\theta}^i)$ with respect to the natural parameter $\bm{\theta}^i$ is
$\bigl\langle(t^i_{y}(X_i) - \eta^i_{y})(t^i_{w}(X_i)-\eta^i_{w})\bigr\rangle_{b_i}$, $\mathcal{J}_{b_i}(\bm{\theta}^i)$ is called a variance matrix and denoted by $\Var_{b_i}[\bm{t}^i(X_i)]$.
Similarly, the weight of degree-two factor node is
$\bigl\langle(t^i_{y_{i,a}}(X_i) - \eta^i_{y_{i,a}})(t^j_{y_{j,a}}(X_j)-\eta^j_{y_{j,a}})\bigr\rangle_{b_a}$.
The corresponding matrix is called a covariance matrix and denoted by $\Cov_{b_a}[\bm{t}^i(X_i), \bm{t}^j(X_j)]$.
Then, from the transfer matrix method, the weight $\mathcal{K}_G(E')$ of a simple generalized loop $E'=\{(i_1, a_1), (i_2, a_1), (i_2, a_2), \dotsc, (i_{\ell}, a_{\ell}), (i_1, a_{\ell})\}\in\mathcal{S}$ is
\begin{align*}
&\tr\Bigl(\Var_{b_{i_1}}[\bm{t}^{i_1}(X_{i_1})]^{-1} \Cov_{b_{(i_1, i_2)}}[\bm{t}^{i_1}(X_{i_1}), \bm{t}^{i_2}(X_{i_2})]\\
&\quad\cdot\Var_{b_{i_2}}[\bm{t}^{i_2}(X_{i_2})]^{-1} \dotsm \Cov_{b_{(i_{\ell}, i_1)}}[\bm{t}^{i_{\ell}}(X_{i_{\ell}}), \bm{t}^{i_1}(X_{i_1})]\Bigr).
\end{align*}
By defining a correlation matrix as
\begin{align*}
&\Cor_{b_a}[\bm{t}^i(X_i), \bm{t}^j(X_j)]:=\Var_{b_i}[\bm{t}^i(X_i)]^{-1/2}\\
&\quad\cdot\Cov_{b_a}[\bm{t}^i(X_i), \bm{t}^j(X_j)]\Var_{b_j}[\bm{t}^j(X_j)]^{-1/2}
\end{align*}
the following lemma is obtained.
\begin{lemma}
The weight $\mathcal{K}_G(E')$ of a simple generalized loop $E'=\{(i_1, a_1), (i_2, a_1), (i_2, a_2), \dotsc, (i_{\ell}, a_{\ell}), (i_1, a_{\ell})\}$ is
\begin{align*}
&\tr\Biggl(\Cor_{b_{a_1}}[\bm{t}^{i_1}(X_{i_1}), \bm{t}^{i_2}(X_{i_2})]\Cor_{b_{a_2}}[\bm{t}^{i_2}(X_{i_2}), \bm{t}^{i_3}(X_{i_3})]\\
&\quad\dotsm \Cor_{b_{a_\ell}}[\bm{t}^{i_{\ell}}(X_{i_{\ell}}), \bm{t}^{i_1}(X_{i_1})]\Biggr).
\end{align*}
\end{lemma}
The following corollary is obtained for a single-cycle factor graph.

\begin{corollary}
For a single-cycle factor graph, it holds
\begin{align*}
&Z(G) = Z_{\mathrm{Bethe}}((b_i)_{i\in V}, (b_a)_{a\in F})\\
&\quad\cdot \Bigl(1 +\tr\Bigl(\Cor_{b_{a_1}}[\bm{t}^{i_1}(X_{i_1}), \bm{t}^{i_2}(X_{i_2})] \\
&\quad\dotsm \Cor_{b_{a_\ell}}[\bm{t}^{i_{\ell}}(X_{i_{\ell}}), \bm{t}^{i_1}(X_{i_1})]\Bigr)
\Bigr).
\end{align*}
\end{corollary}

The transfer matrix method is also useful for the weight of one-dimensional $E'$ in Lemma~\ref{lem:nbmarginal}. 
For a tree factor graph, Theorem~\ref{thm:non-binary} simply means the Bethe approximation is exact.
However, Lemma~\ref{lem:nbmarginal} is useful even for a tree factor graph.

\begin{corollary}[Correlation matrix on a tree factor graph~{\cite[Proposition B.1]{watanabe2010thesis}}]\label{cor:cortree}
For a tree factor graph $G$, the correlation matrix for $i,j\in V$ is decomposed to
\begin{align*}
&\Cor_{p}[\bm{t}^i(X_i), \bm{t}^j(X_j)] =
\Cor_{p}[\bm{t}^i(X_i), \bm{t}^{i_1}(X_{i_1})]\\
&\quad\cdot\Cor_{p}[\bm{t}^{i_1}(X_{i_1}), \bm{t}^{i_2}(X_{i_2})]\dotsm
\Cor_{p}[\bm{t}^{i_\ell}(X_{i_\ell}), \bm{t}^{j}(X_{j})]
\end{align*}
where $(i,i_1\in V,i_2\in V,\dotsc,i_\ell\in V,j)$ is the unique path from $i$ to $j$.
\end{corollary}
\begin{IEEEproof}
When $\ell=0$, i.e., $i$ and $j$ are adjacent, the lemma is trivial.
For $\ell\ge 1$, it is sufficient to prove
\begin{align*}
&\Cov_{p}[\bm{t}^i(X_i), \bm{t}^j(X_j)] =
\Cov_{p}[\bm{t}^i(X_i), \bm{t}^{i_1}(X_{i_1})]\\
&\quad\cdot\Cor_{p}[\bm{t}^{i_1}(X_{i_1}), \bm{t}^{i_2}(X_{i_2})]\\
&\quad\dotsm
\Cor_{p}[\bm{t}^{i_{l-1}}(X_{i_{l-1}}), \bm{t}^{i_\ell}(X_{i_\ell})]
\Cov_{p}[\bm{t}^{i_\ell}(X_{i_\ell}), \bm{t}^{j}(X_{j})].
\end{align*}
Recall that for a tree factor graph, 
the set of pseudo-marginals $((b_i)_{i\in V}, (b_a)_{a\in F})$ on the stationary point of the Bethe free energy is unique
and consists of exact marginal distributions.
Let $C = \{i,j\}$ and
\begin{align*}
g(x_i, x_j) &=
(t^i_{k}(x_i) - \eta^i_{k})
(t^j_{l}(x_j) - \eta^j_{l})\\
&=
\frac{\partial \log b_i(x_i)}{\partial \theta^i_{k}}\frac{\partial \log b_j(x_j)}{\partial \theta^j_{l}}
\end{align*}
for arbitrary fixed $k, l\in\mathcal{X}\setminus\{0\}$ for Lemma~\ref{lem:nbmarginal}.
In this case, $F_C=\varnothing$.
Since $\mathcal{Z}_G(E')=0$ for $E'\subseteq E$ generating degree-one variable node or degree-one factor node except for $i$ and $j$,
$\mathcal{Z}_G(E')$ can be non-zero only for $E'=\varnothing$ and $E'$ being the set of edges in the unique path between $i$ and $j$.
Since $b_C(x_i, x_j)=b_i(x_i)b_j(x_j)$, the weight of the empty set is zero.
The corollary is obtained from Lemmas~\ref{lem:nbmarginal} and \ref{lem:trans} and
\begin{align*}
&\left\langle g(\bm{X}_C) \frac{\partial \log b_i(X_i)}{\partial \eta^i_{y_{i,a}}}\frac{\partial \log b_j(X_j)}{\partial \eta^j_{y_{j,a'}}}\right\rangle_{b_C}\\
&=
\left\langle \frac{\partial \log b_i(X_i)}{\partial \theta^i_{k}}\frac{\partial \log b_i(X_i)}{\partial \eta^i_{y_{i,a}}}\right\rangle_{b_i}\\
&\quad\cdot
\left\langle \frac{\partial \log b_j(X_j)}{\partial \theta^j_{l}}\frac{\partial \log b_j(X_j)}{\partial \eta^j_{y_{j,a'}}}\right\rangle_{b_j}
=\delta(k, y_{i,a}) \delta(l, y_{j,a'}).
\end{align*}
\end{IEEEproof}
Lemma~\ref{lem:nbmarginal} can be also used for generalizing Corollary~\ref{cor:cortree} to correlation among more than two variables 
and to general factor graphs.
Lemma~\ref{lem:nbmarginal} may be also useful for bounding correlations by a sum of weights among all paths like~\cite{PhysRev.162.480} and \cite{von1995taming}.

\section{Loop calculus for continuous alphabets}\label{sec:lcont}
In this section, loop calculus is generalized to continuous alphabets.
A generalization of loop calculus to continuous alphabet was originally obtained by Xiao and Zhou~\cite{xiao2011partition}.
\begin{lemma}\label{lem:lcont}
For any $((b_i)_{i\in V}, (b_a)_{a\in F})\in\mathrm{IS}(\mathcal{F}_{\mathrm{Bethe}})$,
\begin{equation*}
Z(G) = Z_\mathrm{Bethe}((b_i)_{i\in V},(b_a)_{a\in F})\sum_{E'\subseteq E} \bar{\mathcal{K}}_G(E')
\end{equation*}
where
\begin{align*}
&\bar{\mathcal{K}}_G(E') = \int \prod_{a\in F} \left\langle\prod_{i\in\partial a, (i,a)\in E'} \frac{\delta(X_i - v_{i,a})-b_i(v_{i,a})}{\sqrt{b_i(v_{i,a})}}\right\rangle_{b_a}\\
&\cdot \prod_{i\in V}\left\langle\prod_{a\in\partial i, (i,a)\in E'} \frac{\delta(X_i - v_{i,a})-b_i(v_{i,a})}{\sqrt{b_i(v_{i,a})}}\right\rangle_{b_i}
\prod_{(i,a)\in E'} \mathrm{d}v_{i,a}.
\end{align*}
\end{lemma}
The proof of the above lemma is shown in Appendix~\ref{apx:cont}
using Lemma~\ref{lem:xiao}, which was obtained by Xiao and Zhou~\cite{xiao2011partition}.
Lemma~\ref{lem:loopexp} in Appendix~\ref{apx:xiao} for finite alphabets shows
 that the weight in Lemma~\ref{lem:lcont} is equal to the weight in Theorem~\ref{thm:non-binary}.
Similarly, loop calculus for marginal distribution is also obtained.

\begin{lemma}\label{lem:lmcont}
Let $C\subseteq V$ and $g\colon \mathcal{X}^{|C|}\to \mathbb{C}$.
For any $((b_i)_{i\in V}, (b_a)_{a\in F})\in\mathrm{IS}(\mathcal{F}_{\mathrm{Bethe}})$,
\begin{align*}
&Z(G)\langle g(\bm{X}_C)\rangle_p \\
&= Z_\mathrm{Bethe}((b_i)_{i\in V},(b_a)_{a\in F})\sum_{E'\subseteq E\setminus E(F_C)} \bar{\mathcal{K}}^g_G(E')
\end{align*}
where
\begin{align*}
\bar{\mathcal{K}}^g_G(E') &:= \int \prod_{(i,a)\in E'} \mathrm{d}v_{i,a}\\
&\cdot \prod_{a\in F\setminus F_C} \left\langle\prod_{i\in\partial a, (i,a)\in E'} \frac{\delta(X_i - v_{i,a})-b_i(v_{i,a})}{\sqrt{b_i(v_{i,a})}}\right\rangle_{b_a}\\
&\cdot\prod_{i\in V\setminus C}\left\langle\prod_{a\in\partial i, (i,a)\in E'} \frac{\delta(X_i - v_{i,a})-b_i(v_{i,a})}{\sqrt{b_i(v_{i,a})}}\right\rangle_{b_i}\\
&\cdot
\left\langle g(\bm{X}_C)\prod_{a\in\partial i, (i,a)\in E'} \frac{\delta(X_i - v_{i,a})-b_i(v_{i,a})}{\sqrt{b_i(v_{i,a})}}\right\rangle_{b_C}
.
\end{align*}
\end{lemma}
Similarly to Section~\ref{sec:simple} for finite alphabets, the expressions of weights of simple generalized loops can be simplified.

\begin{lemma}\label{lem:lconts}
For a simple generalized loop $E'=\{(i_1, a_1), (i_2, a_1), (i_2, a_2), \dotsc, (i_{\ell}, a_{\ell}), (i_1, a_{\ell})\}$,
it holds
\begin{equation}
\bar{\mathcal{K}}_G(E') 
= \int \frac{b_{i_1,E'}(v, v) - b_{i_1}(v)^2}{b_{i_1}(v)}\mathrm{d}v
\label{eq:Kcont}
\end{equation}
where
\begin{align*}
b_{i_1,E'}(x_{i_1}, x'_{i_1}) &:= \int\prod_{s=2}^\ell\mathrm{d}x_s\\
&\frac{b_{a_1}(x_{i_1}, x_{i_2})\dotsm b_{a_{\ell-1}}(x_{i_{\ell-1}}, x_{i_\ell})b_{a_\ell}(x_{i_\ell}, x'_{i_1})}{\prod_{s=2}^{\ell}b_{i_s}(x_{i_s})}.
\end{align*}
For $C=\{i_1\}$, it holds
\begin{equation}
\bar{\mathcal{K}}^g_G(E')
= \int \frac{b_{i_1,E'}(v, v) - b_{i_1}(v)^2}{b_{i_1}(v)}g(v)\mathrm{d}v.
\label{eq:Kgcont}
\end{equation}
\end{lemma}
\begin{IEEEproof}
The proof is similar to the proof of Lemma~\ref{lem:trans}.
It holds
\begin{align}
&\int\mathrm{d}v_{i,a} \frac{\delta(x_i - v_{i,a})-b_i(v_{i,a})}{\sqrt{b_i(v_{i,a})}}\nonumber\\
&\quad\cdot\left\langle \frac{\left(\delta(X_i - v_{i,a})-b_i(v_{i,a})\right)\left(\delta(X_i - v_{i,a'})-b_i(v_{i,a'})\right)}{\sqrt{b_i(v_{i,a})b_i(v_{i,a'})}}\right\rangle_{b_i}
\nonumber\\
&=
\frac{\delta(x_i - v_{i,a'})-b_i(v_{i,a'})}{\sqrt{b_i(v_{i,a'})}}.
\label{eq:proj}
\end{align}
Hence, from Lemma~\ref{lem:lcont}, it holds
\begin{align}
&\bar{\mathcal{K}}_G(E') 
= \int\prod_{s=1}^\ell\mathrm{d}v_{i_s} \nonumber\\
&\prod_{s=1}^{\ell}
\left\langle \rule{0cm}{0.8cm}\right.\frac{\delta(X_{i_s} - v_{i_s})-b_{i_s}(v_{i_s})}{\sqrt{b_{i_s}(v_{i_s})}}\nonumber\\
&\quad\cdot\frac{\delta(X_{i_{s+1}} - v_{i_{s+1}})-b_{i_{s+1}}(v_{i_{s+1}})}{\sqrt{b_{i_{s+1}}(v_{i_{s+1}})}}\left.\rule{0cm}{0.8cm}\right\rangle_{b_{a_s}} \nonumber\\
&= \int \prod_{s=1}^{\ell} \frac{b_{a_s}(v_{i_s}, v_{i_{s+1}}) - b_{i_s}(v_{i_s})b_{i_{s+1}}(v_{i_{s+1}})}{\sqrt{b_{i_s}(v_{i_s})b_{i_{s+1}}(v_{i_{s+1}})}} \prod_{s=1}^\ell\mathrm{d}v_{i_s}
\label{eq:lsingle}
\end{align}
where $i_{\ell+1}$ is regarded as $i_1$.
Here, it holds
\begin{align}
&\int
\frac{b_{a_s}(v_{i_s}, v_{i_{s+1}}) - b_{i_s}(v_{i_s})b_{i_{s+1}}(v_{i_{s+1}})}{\sqrt{b_{i_s}(v_{i_s})b_{i_{s+1}}(v_{i_{s+1}})}}\nonumber\\
&\cdot \frac{b_{a_{s+1}}(v_{i_{s+1}}, v_{i_{s+2}}) - b_{i_{s+1}}(v_{i_{s+1}})b_{i_{s+2}}(v_{i_{s+2}})}{\sqrt{b_{i_{s+1}}(v_{i_{s+1}})b_{i_{s+2}}(v_{i_{s+2}})}}
\mathrm{d}v_{i_{s+1}}\nonumber\\
&=
\frac{b_{a_{s,s+1}}(v_{i_s}, v_{i_{s+2}}) - b_{i_s}(v_{i_s})b_{i_{s+2}}(v_{i_{s+2}})}{\sqrt{b_{i_s}(v_{i_s})b_{i_{s+2}}(v_{i_{s+2}})}}
\label{eq:rec}
\end{align}
where 
\begin{align}
&b_{a_{s,s+1}}(v_{i_s}, v_{i_{s+2}}) \nonumber\\
&\quad:= 
\int \frac{b_{a_s}(v_{i_s}, v_{i_{s+1}})b_{a_{s+1}}(v_{i_{s+1}}, v_{i_{s+2}})}{b_{i_{s+1}}(v_{i_{s+1}})}\mathrm{d}v_{i_{s+1}}.
\label{eq:ss}
\end{align}
By applying~\eqref{eq:rec} recursively to~\eqref{eq:lsingle}, one obtains \eqref{eq:Kcont}.
The equation~\eqref{eq:Kgcont} is also obtained in the same way.
\end{IEEEproof}

The above lemma can be further simplified for the Gaussian model, which is defined by
\begin{equation*}
p(\bm{x}; G) := \frac1{Z(G)} \exp\left\{-\frac12 \sum_{i,j\in\{1,2,\dotsc,N\}} J_{i,j}x_i x_j + \sum_{i=1}^N h_i x_i\right\}
\end{equation*}
for a positive-definite symmetric matrix $J$, $h\in\mathbb{R}^N$ and $\bm{x}\in\mathbb{R}^N$.
The partition function of the Gaussian model is $\sqrt{\det(J)/(2\pi)^N}$. Similarly, the variance covariance matrix and the expectation of $\bm{X}\in\mathbb{R}^N$ obeying $p$
are $J^{-1}$ and $-J^{-1}h$, respectively.
Although these three quantities can be computed in $O(N^3)$ time by the Gaussian elimination,
it is often desired to approximate them more efficiently.

\begin{lemma}
For the Gaussian model,
a weight of a simple generalized loop $E'=\{(i_1, a_1), (i_2, a_1), (i_2, a_2), \dotsc,\allowbreak (i_{\ell}, a_{\ell}), (i_1, a_{\ell})\}$ in Lemma~\ref{lem:lconts} is
\begin{align}
\bar{\mathcal{K}}_G(E') &= \frac{\Cor_{b_{i_1,E'}}[X_{i_1}, X'_{i_1}]}{1 - \Cor_{b_{i_1,E'}}[X_{i_1}, X'_{i_1}]}\label{eq:Gs}\\
\bar{\mathcal{K}}^g_G(E') &= \frac{1}{1 - \Cor_{b_{i_1,E'}}[X_{i_1}, X'_{i_1}]}\langle g(X_{i_1})\rangle_{\bar{b}_{i_1}} - \langle g(X_{i_1})\rangle_{b_{i_1}}\label{eq:Gsm}
\end{align}
where
\begin{equation*}
\Cor_{b_{i_1,E'}}[X_{i_1}, X'_{i_1}] := \frac{\Cov_{b_{a_1}}[X_{i_1}, X_{i_2}]\dotsm \Cov_{b_{a_\ell}}[X_{i_\ell}, X'_{i_1}]}{\prod_{s=1}^\ell \Var_{b_{i_s}}[X_{i_s}]}
\end{equation*}
and where $\bar{b}_i$ is a Gaussian distribution with the same expectation as $b_i$ and a variance
\begin{equation*}
\Var_{b_{i_1}}[X_{i_1}]\frac{1+\Cor_{b_{i_1, E'}}[X_{i_1}, X'_{i_1}]}{1-\Cor_{b_{i_1, E'}}[X_{i_1}, X'_{i_1}]}.
\end{equation*}
\end{lemma}
\begin{IEEEproof}
For the Gaussian model, the pseudo-marginals at a stationary point are also Gaussian distribution.
The covariance of $(X_{i_s}, X_{i_{s+1}})$ obeying $b_{a_{s,s+1}}$, defined in~\eqref{eq:ss} is
\begin{align*}
&\Cov_{b_{a_{s,s+1}}}[X_{i_s}, X_{i_{s+2}}]\\
&\quad=\frac{\Cov_{b_{a_s}}[X_{i_s}, X_{i_{s+1}}] \Cov_{b_{a_{s+1}}}[X_{i_{s+1}}, X_{i_{s+2}}]}{\Var_{b_{i_{s+1}}}[X_{i_{s+1}}]}.
\end{align*}
Hence, $b_{i_1,E'}(x_{i_1}, x'_{i_1})$ in~\eqref{eq:Kcont} is a Gaussian distribution with a covariance
\begin{equation*}
\Cov_{b_{i_1,E'}}[X_{i_1}, X'_{i_1}] = \frac{\Cov_{b_{a_1}}[X_{i_1}, X_{i_2}]\dotsm \Cov_{b_{a_\ell}}[X_{i_\ell}, X'_{i_1}]}{\prod_{s=2}^\ell \Var_{b_{i_s}}[X_{i_s}]}
\end{equation*}
and variances $\Var_{b_{i_1,E'}}[X_{i_1}] = \Var_{b_{i_1,E'}}[X'_{i_1}] = \Var_{b_{i_1}}[X_{i_1}]$.
Finally,~\eqref{eq:Gs} is obtained from
\begin{equation*}
\int \frac{b_{i_1,E'}(v, v) - b_{i_1}(v)^2}{b_{i_1}(v)}\mathrm{d}v
=\frac{c}{1-c}
\end{equation*}
where $c:=\Cov_{b_{i_1,E'}}[X_{i_1}, X'_{i_1}]/\Var_{b_{i_1}}[X_{i_1}]$.
The equation~\eqref{eq:Gsm} is also obtained in a similar way.
\end{IEEEproof}

\begin{corollary}\label{cor:singleG}
For a single-cycle factor graph, it holds
\begin{align}
Z(G) &= \frac{Z_{\mathrm{Bethe}}((b_i)_{i\in V},(b_a)_{a\in F})}{1-\Cor_{b_{i_1}, E'}[X_{i_1}, X'_{i_1}]}\nonumber\\
\langle g(X_{i_1})\rangle_p &=
\langle g(X_{i_1})\rangle_{\bar{b}_{i_1}}.
\label{eq:Gsv}
\end{align}
\end{corollary}
From~\eqref{eq:Gsv}, $b_i$ at the stationary point of the Bethe free energy has the correct mean $\langle X\rangle_p$ for single-cycle factor graphs.
Indeed, the fixed point of BP gives the exact means for any Gaussian model~\cite{weiss2001correctness}.

\begin{remark}[Verification by the walk-sum formula]
For variances obtained by the Bethe approximation, from~\eqref{eq:Gsv}, it holds
\begin{align}
&\Var_p[X_{i_1}] = \Var_{b_{i_1}}[X_{i_1}] \frac{1+\Cor_{b_{i_1, E'}}[X_{i_1}, X'_{i_1}]}{1-\Cor_{b_{i_1, E'}}[X_{i_1}, X'_{i_1}]}\nonumber\\
&= \Var_{b_{i_1}}[X_{i_1}]  + 2\Var_{b_{i_1}}[X_{i_1}]\frac{\Cor_{b_{i_1, E'}}[X_{i_1}, X'_{i_1}]}{1-\Cor_{b_{i_1, E'}}[X_{i_1}, X'_{i_1}]}
\label{eq:GW}
\end{align}
for a single-cycle factor graph.
In this remark, 
this equation is confirmed by the walk-sum formula~\cite{malioutov2006walk}.
This remark is less self-contained. See also~\cite{malioutov2006walk} for details.
It holds
\begin{equation*}
J= \sqrt{D} (I- W) \sqrt{D}
\end{equation*}
where $D$ is a diagonal matrix whose $(i,i)$-element is $J_{i,i}$ and where $W=I-D^{-1/2}JD^{-1/2}$ is a symmetric matrix with zero diagonal.
If the spectral radius of $W$ is smaller than 1, it holds
\begin{equation*}
J^{-1} = D^{-1/2}\left(I+W+W^2+\dotsb\right)D^{-1/2}.
\end{equation*}
Hence, if $J$ is walk-summable, i.e., the spectral radius of $|W|$, in which elements are replaced by their absolute value, is smaller than 1 (See~\cite{malioutov2006walk} for details),
it holds
\begin{align*}
\Var_p[X_i] &= \frac1{J_{i,i}}\sum_{w\colon i\xrightarrow[G]{} i} \phi(w)\\
\Cov_p[X_i, X_j] &= \frac1{\sqrt{J_{i,i}J_{j,j}}}\sum_{w\colon i\xrightarrow[G]{} j} \phi(w)
\end{align*}
where $w\colon i\xrightarrow[G]{} j$ denotes a walk from $i$ to $j$ on $G$ and where $\phi(w):= W_{i,i_1}W_{i_1,i_2}\dotsm W_{i_\ell, j}$ for a walk $w=(i, i_1, i_2,\dotsc, i_\ell, j)$.
Since BP is an exact algorithm on the computation tree, it holds
\begin{align*}
\Var_{b_i}[X_i] &= \frac1{J_{i,i}}\sum_{w\colon i\xrightarrow[T_i]{} i} \phi(w)\\
\Cov_{b_a}[X_i, X_j] &= \frac1{\sqrt{J_{i,i}J_{j,j}}}\sum_{w\colon i\xrightarrow[T_{i,j}]{} j} \phi(w)
\end{align*}
where $w\colon i\xrightarrow[T_i]{} i$ is a walk from $i$ to $i$, both being the root variable on the computation tree $T_i$ for $i\in V$
and where $w\colon i\xrightarrow[T_{i,j}]{} j$ is a walk from $i$ to $j$, both on the root edge on the computation tree $T_{i,j}$ for $(i,j)\in V^2$.
Any walk from $i$ on $G$ can be naturally identified with a walk from $i$ on the computation tree $T_i$.
Hence, for confirming~\eqref{eq:GW} we should verify
\begin{align}
&\frac1{J_{i_1,i_1}}\sum_{\substack{w\colon i_1\xrightarrow[G]{} i_1\\ w\colon \text{not } i_1\xrightarrow[T_{i_1}]{} i_1}} \phi(w)\nonumber\\
&=
2\Var_{b_{i_1}}[X_{i_1}]\frac{\Cor_{b_{i_1, E'}}[X_{i_1}, X'_{i_1}]}{1-\Cor_{b_{i_1, E'}}[X_{i_1}, X'_{i_1}]}.
\label{eq:walk-loop}
\end{align}
From
\begin{align*}
\frac{\Cov_{b_a}[X_i, X_j]}{\Var_{b_i}[X_i]} &= \sqrt{\frac{J_{i,i}}{J_{j,j}}}\sum_{w\colon i \xrightarrow[T_{i,j}]{\setminus i} j} \phi(w)
\end{align*}
where $w\colon i\xrightarrow[T_{i,j}]{\setminus i} j$ is a walk from $i$ to $j$ which does not visit $i$ except as initial place, it holds
\begin{align*}
\Cor_{b_{i_1, E'}}[X_{i_1}, X'_{i_1}] &=
\sum_{w\colon i_1\xrightarrow[T_i]{\setminus i_1} i_1'} \phi(w)
\end{align*}
and hence
\begin{align*}
\Var_{b_{i_1}}[X_{i_1}]\Cor_{b_{i_1, E'}}[X_{i_1}, X'_{i_1}] &=
\frac1{J_{i_1,i_1}}\sum_{w\colon i_1\xrightarrow[T_i]{} i_1'} \phi(w)
\end{align*}
which express the weight of walks from $i_1$ to $i'_1$ which is one of the nearest copies of $i_1$ in the computation tree $T_{i_1}$.
The weight of walks from $i_1$ to secondary nearest copies of $i_1$ is 
$2\Var_{b_{i_1}}[X_{i_1}]\Cor_{b_{i_1, E'}}[X_{i_1}, X'_{i_1}]^2$.
Hence, finally, one obtains
\begin{align*}
&\frac1{J_{i_1,i_1}}\sum_{\substack{w\colon i_1\xrightarrow[G]{} i_1\\ w\colon \text{not } i_1\xrightarrow[T_{i_1}]{} i_1}} \phi(w)\\
&=
2\Var_{b_{i_1}}[X_{i_1}]\Bigl(\Cor_{b_{i_1, E'}}[X_{i_1}, X'_{i_1}]+\Cor_{b_{i_1, E'}}[X_{i_1}, X'_{i_1}]^2\\
&\quad +\Cor_{b_{i_1, E'}}[X_{i_1}, X'_{i_1}]^3+\dotsb\Bigr).
\end{align*}
Hence,~\eqref{eq:walk-loop} is verified.
\end{remark}
For the Gaussian model, the multiplicative error in the Bethe approximation is expressed by an infinite product in~\cite{Johnson:2009:ORC}.
The loop calculus for the Fermion model whose partition function can express a determinant of any square matrix is obtained in~\cite{1742-5468-2008-12-P12011}.

\section{Bethe approximation and loop calculus for weighted graph coloring on a regular graph}\label{sec:color}
In this section, simple examples of loop calculus for non-binary graphical models are shown.
The partition function for weighted graph $q$-coloring problem for a parameter $w>0$ is
\begin{equation*}
Z(G) = \sum_{\bm{x}\in\mathcal{X}^N} \prod_{i\in V} w^{\delta(x_i, \underline{0})} \prod_{a\in F} (1-\delta(x_{i_1(a)}, x_{i_2(a)}))
\end{equation*}
where $\{i_1(a), i_2(a)\} := \partial a$ for $a\in F$.
The fixed point equation of belief propagation for the weighted graph $q$-coloring is
\begin{align*}
m_{i\to a}(x) &\propto w^{\delta(x,\underline{0})}\prod_{a'\in\partial i\setminus\{a\}} m_{a' \to i}(x)\\
m_{a\to i}(x) &= \frac1{q-1}\left(1-m_{j\to a}(x)\right).
\end{align*}
Assume that $w=1$ and that the messages are common for all edges.
Then, obviously the uniform distribution is the unique fixed point of the belief propagation.
In that case, by applying Theorem~\ref{thm:non-binary} for $(t_y(x)=\delta(x,y))_{y\in\mathcal{X}\setminus\{\underline{0}\}}$,
one obtains
\begin{align*}
Z(G) &= q^N\left(1-\frac1q\right)^{|E|}\sum_{E'\subseteq E} \sum_{\bm{y}\in(\mathcal{X}\setminus\{\underline{0}\})^{|E'|}}\\
&\quad\prod_{a\in F'}\left[(1-\delta(y_{i_1(a)}, y_{i_2(a)}))\frac1{q(q-1)} - \frac1{q^2}\right]\\
&\quad\cdot\prod_{i\in V}\left[q^{d_i(E')-1}((-1)^{d_i}+\delta(\bm{y}_{\partial i}))\right].
\end{align*}
The above equation can be also obtained from the well-known high temperature expansion for the Potts model~\cite{wu1982potts}.
For $w\ne 1$, if a graph is $k$-regular, we can also assume that all messages are common although such messages does not necessarily minimize
the Bethe free energy.
On the assumption, the normalized Bethe free energy is equivalent to the annealed free energy for random regular graph, i.e.,
$(1/N)\mathcal{F}_{\mathrm{Bethe}} = -\lim_{N\to\infty}(1/N)\log\mathbb{E}[Z]$~\cite{mori2011connection}.
Hence, the Bethe approximation only depends on the size and independent of the connections of edges.
On the assumption of the common messages, all probabilities of nonzero element must be the same at the fixed point.
Hence, the messages are restricted in the exponential family with a single parameter for the sufficient statistic $t(x)=\delta(x,\underline{0})$.
The fixed point equation is
\begin{align*}
\theta_{\mathrm{v}\to \mathrm{f}} &= \log w + (k-1)\theta\left(\eta_{\mathrm{f}\to \mathrm{v}}\right)\\
\eta_{\mathrm{f}\to \mathrm{v}} &= \frac1{q-1}\left(1-\eta\left(\theta_{\mathrm{v}\to\mathrm{f}}\right)\right)
\end{align*}
where
\begin{align*}
\eta(\theta)&:=
\frac{\exp\{\theta\}}{q-1+\exp\{\theta\}},&
\theta(\eta)&:=\log\frac{(q-1)\eta}{1-\eta}.
\end{align*}
Note that for large $w$, the unique fixed point of the above equations is not stable with respect to the forward substitution of
the above equations.
In that case, the backward substitution yields the convergence to the unique fixed point.

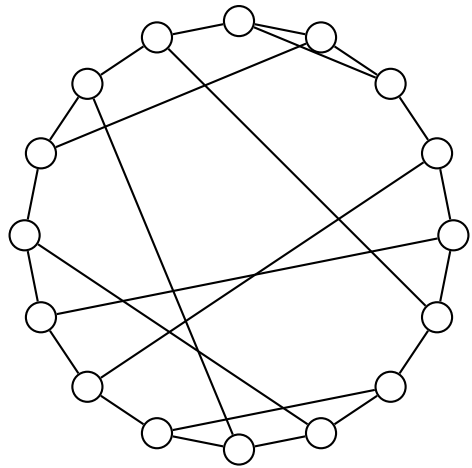
\begin{figure}[t]
\centering
\begin{tikzpicture}
[yshift=20pt, scale=0.57, inner sep=0mm, C/.style={minimum size=4mm,circle,draw=black,thick},
S/.style={minimum size=4mm,rectangle,draw=black,thick}, label distance=1mm]
\node (0) at ({5*cos(0)}, {5*sin(0)}) [C] {};
\node (4) at ({5*cos(22.5)}, {5*sin(22.5)}) [C] {};
\node (15) at ({5*cos(45)}, {5*sin(45)}) [C] {};
\node (6) at ({5*cos(67.5)}, {5*sin(67.5)}) [C] {};
\node (7) at ({5*cos(90)}, {5*sin(90)}) [C] {};
\node (8) at ({5*cos(112.5)}, {5*sin(112.5)}) [C] {};
\node (11) at ({5*cos(135)}, {5*sin(135)}) [C] {};
\node (5) at ({5*cos(157.5)}, {5*sin(157.5)}) [C] {};
\node (1) at ({5*cos(180)}, {5*sin(180)}) [C] {};
\node (9) at ({5*cos(202.5)}, {5*sin(202.5)}) [C] {};
\node (3) at ({5*cos(225)}, {5*sin(225)}) [C] {};
\node (13) at ({5*cos(247.5)}, {5*sin(247.5)}) [C] {};
\node (2) at ({5*cos(270)}, {5*sin(270)}) [C] {};
\node (10) at ({5*cos(292.5)}, {5*sin(292.5)}) [C] {};
\node (14) at ({5*cos(315)}, {5*sin(315)}) [C] {};
\node (12) at ({5*cos(337.5)}, {5*sin(337.5)}) [C] {};
\draw (1) to (5) [thick];
\draw (14) to (13) [thick];
\draw (11) to (2) [thick];
\draw (3) to (13) [thick];
\draw (7) to (15) [thick];
\draw (11) to (5) [thick];
\draw (6) to (5) [thick];
\draw (11) to (8) [thick];
\draw (10) to (14) [thick];
\draw (9) to (3) [thick];
\draw (6) to (15) [thick];
\draw (14) to (12) [thick];
\draw (4) to (15) [thick];
\draw (0) to (4) [thick];
\draw (2) to (10) [thick];
\draw (1) to (9) [thick];
\draw (8) to (7) [thick];
\draw (6) to (7) [thick];
\draw (8) to (12) [thick];
\draw (12) to (0) [thick];
\draw (13) to (2) [thick];
\draw (10) to (1) [thick];
\draw (3) to (4) [thick];
\draw (0) to (9) [thick];
\end{tikzpicture}
\caption{A 3-regular graph of size 16.}
\label{fig:graph}
\end{figure}

Results of numerical calculation for the graph in Fig.~\ref{fig:graph} of size 16, which is generated randomly, is shown in Table~\ref{tbl:color}.
The approximations $Z_{\mathrm{Bethe+loops}}$ and $Z_{\mathrm{Bethe\times loops}}$ are defined as
\begin{align*}
Z_{\mathrm{Bethe+loops}} &:= Z_{\mathrm{Bethe}}\left(1+\sum_{E'\in\mathcal{S}}\mathcal{K}_G(E')\right)\\
Z_{\mathrm{Bethe\times loops}} &:= Z_{\mathrm{Bethe}}\prod_{E'\in\mathcal{S}}\left(1+\mathcal{K}_G(E')\right).
\end{align*}
In Table~\ref{tbl:color}, except for the cases $q=3$, the approximations using loop corrections
are better than the original Bethe approximation.
The new approximations are accurate especially for large $q$.
For the case $w=1.5$, $q=3$, the fixed point is unstable with respect to the forward substitution.
The approximation using the edge zeta function suggested in~\cite{mori2012new} is similar to $Z_{\mathrm{Bethe\times loops}}$,
and can be efficiently computed.
It may also give an efficient and accurate approximation.

\begin{table*}[t]
\renewcommand{\arraystretch}{1.3}
\caption{Error of the Bethe approximation with loop corrections for the 3-regular graph in Fig.~\ref{fig:graph}}
\label{tbl:color}
\centering
\begin{tabular}{c|c|c|c|c}
\hline
& $Z(G)$ & $Z_{\mathrm{Bethe}}/Z(G)$ & $Z_{\mathrm{Bethe+loops}}/Z(G)$ & $Z_{\mathrm{Bethe\times loops}}/Z(G)$\\
\hline
$w=1$, $q=3$ & $2628$& $0.973$ & $1.117$ & $1.060$\\
\hline
$w=1$, $q=4$ & $4143720$ & $1.040$ & $1.011$ & $1.003$\\
\hline
$w=1$, $q=9$ & $108384232602240$ & $1.012$ & $1.00007$ & $1.00001$\\
\hline \hline
$w=1.5$, $q=3$ & $25035.75$& $0.952$ & $1.130$ & $1.070$\\
\hline
$w=1.5$, $q=4$ & $23205262.5$& $1.035$ & $1.013$ & $1.004$\\
\hline
$w=1.5$, $q=9$ & $244818663513163.34$& $1.013$ & $1.00008$ & $1.00002$\\
\end{tabular}
\end{table*}

\section*{Acknowledgment}
The author would like to thank Prof.\ Toshiyuki~Tanaka for many discussions.
The author would like to thank Dr.\ Pascal O.~Vontobel for many discussions and especially for giving a motivation for Lemma~\ref{lem:diag}.
The author would like to thank the editor and the anonymous reviewers for their many insightful comments and suggestions.

\appendices

\section{Exponential family}
\label{apx:ef}

\begin{definition}[Exponential family]
The exponential family is a parametric family of probability measures.
Let $\Theta\subseteq\mathbb{R}^d$ be a space of parameter and $\mathcal{X}$ be a sample space. Then, 
probability mass function (or probability density function) of exponential family is expressed as
\begin{equation*}
p(x;\bm{\theta}):= \frac1{Z(\bm{\theta})} \exp\left\{\sum_{k=1}^d \theta_k t_k(x)\right\}
\end{equation*}
for $\bm{\theta}\in\Theta$
using a set of functions $(t_k\colon \mathcal{X}\to\mathbb{R})_{k=1,\dots,d}$ called a sufficient statistic
where
\begin{equation*}
Z(\bm{\theta}) := \sum_{x\in\mathcal{X}}\exp\left\{\sum_{k=1}^d \theta_k t_k(x)\right\}.
\end{equation*}
Here, the parameter $\bm{\theta}$ is called a natural parameter.
\end{definition}
The map $\theta \mapsto p(x;\bm{\theta})$ is injection if and only if functions $t_1(x), \dotsc, t_d(x),$ and $1$ are linearly independent.
In this paper, when we deal with exponential families, the linear independence is always assumed.
For an exponential family, there is a dual parameter $\bm{\eta}=(\eta_k := \langle t_k(X)\rangle_p)_{k=1,\dotsc,d}$ called
an expectation parameter.
\begin{example}[Distribution on a finite set]
The family of distributions on a finite set $\mathcal{X}=\{\underline{0},\underline{1},\dotsc,\underline{q-1}\}$ can be regarded as
$(q-1)$-dimensional exponential family with a sufficient statistic $(t_{x}(x')=\delta(x,x'))_{x\in\mathcal{X}\setminus \{\underline{0}\}}$.
In this case, 
$\theta_x=\log [p(x\mid \bm{\theta})/ p(\underline{0}\mid\bm{\theta})]$ and
$\eta_x=p(x\mid \bm{\theta})$ for $x\in\mathcal{X}\setminus \{\underline{0}\}$.
\end{example}
Let $H:=\{\langle t_k(X)\rangle_{p(x;\bm{\theta})}, \theta\in\Theta\}$ be the space of exponential parameter.
For each natural parameter $\bm{\theta}$ there exists corresponding expectation parameter $\bm{\eta}=\bm{\eta}(\bm{\theta})$.
The function $\bm{\eta}(\bm{\theta})\colon\Theta\to H$ can be explicitly expressed as
\begin{equation}
\bm{\eta}(\bm{\theta})_k=
\frac{\partial \log Z(\bm{\theta})}{\partial \theta_k}.
\label{eq:e0}
\end{equation}
The function $\log Z(\bm{\theta})$ is strictly convex since
\begin{equation*}
\frac{\partial^2 \log Z(\bm{\theta})}{\partial \theta_k\partial \theta_l}=\langle t_k(x)t_l(x)\rangle_p - \eta_k\eta_l
\end{equation*}
and since sufficient statistics are linearly independent.
As consequence, the map $\bm{\eta}(\bm{\theta})$ is injection since if $\bm{\eta}(\bm{\theta})=\bm{\eta}(\bm{\theta}')$ then $\bm{\theta}=\bm{\theta}'$ due to
the strict convexity of $\log Z(\bm{\theta})$.
Hence, there is a one-to-one correspondence between $\bm{\theta}$ and $\bm{\eta}$.
From this view, we regard the parameters $\bm{\theta}$ and $\bm{\eta}$ as coordinate systems.
The inverse function of $\bm{\eta}(\bm{\theta})$ is denoted by $\bm{\theta}(\bm{\eta})$.
The probability mass function (probability density function) $p$ can be regarded as a function of $\bm{\eta}$
and denoted by $p(x;\bm{\eta}):=p(x;\bm{\theta}(\bm{\eta}))$.
From~\eqref{eq:e0}, $\bm{\theta}(\bm{\eta})\colon H\to\Theta$ can be expressed as
\begin{equation*}
\bm{\theta}(\bm{\eta})=
\mathop{\mathrm{argmax}}_{\bm{\theta}}\left\{\sum_{k=1}^d\theta_k\eta_k-\log Z(\bm{\theta})\right\}.
\end{equation*}
Let
\begin{equation*}
\varphi(\bm{\eta}):=
\max_{\bm{\theta}}\left\{\sum_{k=1}^d\theta_k\eta_k-\log Z(\bm{\theta})\right\}=\langle\log p(X;\bm{\eta})\rangle_{p(x;\bm{\eta})}.
\end{equation*}
Then, it holds
\begin{align*}
&\frac{\partial \varphi(\bm{\eta})}{\partial \eta_k}
=
\frac{\partial \sum_{l=1}^d \bm{\theta}(\bm{\eta})_l \eta_l-\log Z(\bm{\theta}(\bm{\eta}))}{\partial \eta_k}\\
&=
\sum_{l=1}^d \frac{\partial\bm{\theta}(\bm{\eta})_l}{\partial \eta_k}\eta_l + \bm{\theta}(\bm{\eta})_k
-\sum_{l=1}^d \left.\frac{\partial\log Z(\bm{\theta})}{\partial \theta_l}\right|_{\bm{\theta}=\bm{\theta}(\bm{\eta})}\frac{\partial \bm{\theta}(\bm{\eta})_l}{\partial \eta_k}\\
&=\bm{\theta}(\bm{\eta})_k.
\end{align*}
Since the Hessian matrix of $\varphi(\bm{\eta})$ is also Jacobian matrix $\frac{\mathrm{d} \bm{\theta}(\bm{\eta})}{\mathrm{d}\bm{\eta}}$,
which is the inverse matrix of $\frac{\mathrm{d} \bm{\eta}(\bm{\theta})}{\mathrm{d}\bm{\theta}}$, the Hessian matrix of $\varphi(\bm{\eta})$ is positive-definite,
and hence $\varphi(\bm{\eta})$ is strictly convex.
Hence, it also holds
\begin{equation*}
\bm{\eta}(\bm{\theta})=
\mathop{\mathrm{argmax}}_{\bm{\eta}}\left\{\sum_{k=1}^d\theta_k\eta_k-\varphi(\bm{\eta})\right\}
\end{equation*}
and
\begin{equation*}
\psi(\bm{\theta}):=
\max_{\bm{\eta}}\left\{\sum_{k=1}^d\theta_k\eta_k-\varphi(\bm{\eta})\right\}
=\log Z(\bm{\theta})
\end{equation*}
In information geometry, the coordinate systems $\bm{\theta}$ and $\bm{\eta}$ are said to be dual and are known to satisfy
\begin{equation*}
\left\langle \frac{\partial\log p(X;\bm{\theta})}{\partial \theta_k}\frac{\partial\log p(X;\bm{\eta})}{\partial \eta_l}\right\rangle_{p} = \delta(k, l).
\end{equation*}
The above equality is easily confirmed via
\begin{align}
&\left\langle \frac{\partial\log p(X;\bm{\theta})}{\partial \theta_k}\frac{\partial\log p(X;\bm{\eta})}{\partial \eta_l}\right\rangle_{p} \nonumber\\
&= 
\left\langle \left(t_k(X) - \eta_k\right)\frac1{p(X;\bm{\eta})}\frac{\partial p(X;\bm{\eta})}{\partial \eta_l}\right\rangle_{p}
 = \frac{\partial \eta_k}{\partial \eta_l}.
\label{eq:dual}
\end{align}
The Hessian matrix $\mathcal{J}(\bm{\theta})$ of $\psi(\bm{\theta})$ is called the Fisher information matrix with respect to the natural parameter, whose $(k,l)$-element is
\begin{equation*}
\mathcal{J}_p(\bm{\theta})_{k,l}=
\left\langle \frac{\partial\log p(X;\bm{\theta})}{\partial \theta_k}\frac{\partial\log p(X;\bm{\theta})}{\partial \theta_l}\right\rangle_{p}.
\end{equation*}
Similarly, the Fisher information matrix $\mathcal{J}(\bm{\eta})$ with respect to the expectation parameter is defined as
\begin{equation*}
\mathcal{J}_p(\bm{\eta})_{k,l}=
\left\langle \frac{\partial\log p(X;\bm{\eta})}{\partial \eta_k}\frac{\partial\log p(X;\bm{\eta})}{\partial \eta_l}\right\rangle_{p}.
\end{equation*}
The Fisher information matrix $\mathcal{J}(\bm{\eta})$ with respect to the expectation parameter is the Hessian matrix of $\varphi(\bm{\eta})$ and
the inverse matrix of the Fisher information matrix $\mathcal{J}(\bm{\theta})$ with respect to the natural parameter.

\section{Proof of Lemma~\ref{lem:lcont}}\label{apx:cont}
The following lemma was obtained by Xiao and Zhou.
\begin{lemma}[\cite{xiao2011partition}]\label{lem:xiao}
For any $((b_i)_{i\in V}, (b_a)_{a\in F})\in\mathrm{IS}(\mathcal{F}_{\mathrm{Bethe}})$,
\begin{equation*}
Z(G) = Z_\mathrm{Bethe}((b_i)_{i\in V},(b_a)_{a\in F})\sum_{E'\subseteq E} \bar{\mathcal{K}}_G(E')
\end{equation*}
where
\begin{align*}
\bar{\mathcal{K}}_G(E') &:= \int \prod_{(i,a)\in E}\mathrm{d}v_{i,a}\prod_{a\in F} b_a(\bm{v}_{\partial a, a}) \\
&\prod_{i\in V}\left\langle\prod_{a\in\partial i, (i,a)\in E'} \frac{\delta(X_i-v_{i,a})-b_i(v_{i,a})}{b_i(v_{i,a})}\right\rangle_{b_i}.
\end{align*}
\end{lemma}
\begin{IEEEproof}
The ratio $Z(G)/ Z_{\mathrm{Bethe}}((b_i)_{i\in V},(b_a)_{a\in F})$ is equal to
\begin{align*}
&\int \prod_{a\in F}\frac{b_a(\bm{x}_{\partial a})}{\prod_{i\in\partial a}b_i(x_i)}
\prod_{i\in V}b_i(x_i) \prod_{i\in V}\mathrm{d}x_{i}\\
&=
\int\prod_{i\in V}\mathrm{d}x_{i}\prod_{(i,a)\in E}\mathrm{d}v_{i,a}
\prod_{a\in F}b_a(\bm{v}_{\partial a, a})
\prod_{i\in V}b_i(x_i)\\
&\quad\cdot \prod_{(i,a)\in E} \frac{\delta(v_{i,a}- x_i)}{b_i(v_{i,a})}\\
&=
\int\prod_{i\in V}\mathrm{d}x_{i}\prod_{(i,a)\in E}\mathrm{d}v_{i,a}
\prod_{a\in F}b_a(\bm{v}_{\partial a, a})
\prod_{i\in V}b_i(x_i)\\
&\quad\cdot\prod_{(i,a)\in E} \left[1+\frac{\delta(v_{i,a}- x_i) - b_i(v_{i,a})}{b_i(v_{i,a})}\right]\\
&=
\int\prod_{i\in V}\mathrm{d}x_{i}\prod_{(i,a)\in E}\mathrm{d}v_{i,a} \prod_{a\in F}b_a(\bm{v}_{\partial a, a})
\prod_{i\in V}b_i(x_i)\\
&\quad\cdot\sum_{E'\subseteq E} \prod_{(i,a)\in E'} \frac{\delta(v_{i,a}- x_i) - b_i(v_{i,a})}{b_i(v_{i,a})}\\
&=
\sum_{E'\subseteq E}
\int\prod_{(i,a)\in E}\mathrm{d}v_{i,a} \prod_{a\in F}b_a(\bm{v}_{\partial a, a})\\
&\quad\cdot\prod_{i\in V} \left\langle\prod_{a\in\partial i, (i,a)\in E'} \frac{\delta(v_{i,a}- X_i) - b_i(v_{i,a})}{b_i(v_{i,a})}\right\rangle_{b_i}.
\end{align*}
\end{IEEEproof}
\begin{IEEEproof}[Proof of Lemma~\ref{lem:lcont}]
The weight $\bar{\mathcal{K}}_G(E')$ is equal to
\begin{align*}
&\int \prod_{(i,a)\in E}\mathrm{d}v_{i,a}\prod_{a\in F}b_a(\bm{v}_{\partial a, a})\\
&\quad\cdot \prod_{i\in V} \left\langle\prod_{a\in\partial i, (i,a)\in E'} \frac{\delta(X_i-v_{i,a}) - b_i(v_{i,a})}{b_i(v_{i,a})}\right\rangle_{b_i}\\
&=
\int\prod_{(i,a)\in E}\mathrm{d}v_{i,a}\prod_{(i,a)\in E}\mathrm{d}w_{i,a}\prod_{(i,a)\in E} \delta(w_{i,a}-v_{i,a}) \\
&\quad\cdot \prod_{a\in F}b_a(\bm{w}_{\partial a, a})\prod_{i\in V} \left\langle\prod_{a\in\partial i, (i,a)\in E'} \frac{\delta(X_i-v_{i,a}) - b_i(v_{i,a})}{b_i(v_{i,a})}\right\rangle_{b_i}
\\
&=
\int \prod_{(i,a)\in E}\mathrm{d}v_{i,a}\prod_{(i,a)\in E}\mathrm{d}w_{i,a}\\
&\quad\cdot\prod_{a\in F}b_a(\bm{w}_{\partial a, a})
\prod_{i\in V} \left\langle\prod_{a\in\partial i, (i,a)\in E'} \frac{\delta(X_i-v_{i,a}) - b_i(v_{i,a})}{b_i(v_{i,a})}\right\rangle_{b_i}\\
&\quad\cdot
\prod_{(i,a)\in E} b_i(v_{i,a})
\prod_{(i,a)\in E} \left[1+\frac{\delta(w_{i,a}-v_{i,a})-b_i(v_{i,a})}{b_i(v_{i,a})}\right]\\
&=
\int \prod_{(i,a)\in E}\mathrm{d}v_{i,a}\prod_{(i,a)\in E}\mathrm{d}w_{i,a}\\
&\quad\cdot\prod_{a\in F}b_a(\bm{w}_{\partial a, a})
\prod_{i\in V} \left\langle\prod_{a\in\partial i, (i,a)\in E'} \frac{\delta(X_i-v_{i,a}) - b_i(v_{i,a})}{b_i(v_{i,a})}\right\rangle_{b_i}\\
&\quad\cdot \prod_{(i,a)\in E} b_i(v_{i,a})
\sum_{E''\subseteq E} \prod_{(i,a)\in E''} \frac{\delta(w_{i,a}-v_{i,a})-b_i(v_{i,a})}{b_i(v_{i,a})}\\
&=
\sum_{E''\subseteq E} \int \prod_{(i,a)\in E}\mathrm{d}v_{i,a}\prod_{(i,a)\in E} b_i(v_{i,a})\\
&\quad\cdot\prod_{a\in F}\left\langle\prod_{i\in\partial a, (i,a)\in E''}\frac{\delta(X_i-v_{i,a})-b_i(v_{i,a})}{b_i(v_{i,a})}\right\rangle_{b_a}\\
&\quad\cdot \prod_{i\in V} \left\langle\prod_{a\in\partial i, (i,a)\in E'} \frac{\delta(X_i-v_{i,a}) - b_i(v_{i,a})}{b_i(v_{i,a})}\right\rangle_{b_i}.
\end{align*}
In the last equation, the terms corresponding to $E''\ne E'$ are zero.
\end{IEEEproof}
Lemma~\ref{lem:lmcont} is also proved in similar way as the above derivation and Lemma~\ref{lem:bmarginal}.

\section{Relationship with Xiao and Zhou's loop calculus for continuous alphabets}\label{apx:xiao}
The weight~\eqref{eq:nbloop} of generalized loop for non-binary finite alphabets has the following form.
\begin{lemma}\label{lem:loopexp}
\begin{align*}
\mathcal{K}_G(E')&=\sum_{\bm{z}_{E'}\in\mathcal{X}^{|E'|}}
\prod_{a\in F}
\left\langle \prod_{i\in\partial a, (i,a)\in E'} \frac{\delta\left(z_{i,a}, X_i\right) - b_i(z_{i,a})}{\sqrt{b_i(z_{i,a})}}\right\rangle_{b_a}\\
&\quad\cdot\prod_{i\in V}
\left\langle \prod_{a\in\partial i, (i,a)\in E'} \frac{\delta\left(z_{i,a}, X_i\right) - b_i(z_{i,a})}{\sqrt{b_i(z_{i,a})}}\right\rangle_{b_i}.
\end{align*}
\end{lemma}
\begin{IEEEproof}
Once one has
\begin{align}
&\mathcal{K}_G(E')=\sum_{\bm{z}\in\mathcal{X}^{2|E'|}}\nonumber\\
&\quad\prod_{a\in F}
\left\langle \prod_{i\in\partial a, (i,a)\in E'} \frac{\delta\left(z_{(i,a),a}, X_i\right) - b_i(z_{(i,a),a})}{\sqrt{b_i(z_{(i,a),a})}}\right\rangle_{b_a}\nonumber\\
&\quad\cdot \prod_{i\in V}
\left\langle \prod_{a\in\partial i, (i,a)\in E'} \frac{\delta\left(z_{(i,a),i}, X_i\right) - b_i(z_{(i,a),i})}{\sqrt{b_i(z_{(i,a),i})}}\right\rangle_{b_i}\nonumber\\
&\quad\cdot \prod_{(i,a)\in E'}
\left\langle \rule{0cm}{0.6cm}\right.\frac{\delta\left(z_{(i,a),a}, X_i\right) - b_i(z_{(i,a),a})}{\sqrt{b_i(z_{(i,a),a})}}\nonumber\\
&\quad\cdot\frac{\delta\left(z_{(i,a),i}, X_i\right) - b_i(z_{(i,a),i})}
{\sqrt{b_i(z_{(i,a),i})}}\left.\rule{0cm}{0.6cm}\right\rangle_{b_i}\label{eq:contr}
\end{align}
then the lemma is obtained by~\eqref{eq:contr} and \eqref{eq:proj}. 
The equality~\eqref{eq:contr} can be proved by using the Holant theorem~\eqref{eq:holant} from~\eqref{eq:nbloop} using the alphabet $\mathcal{X}\setminus\{\underline{0}\}$ to different representation using the alphabet $\mathcal{X}$.
Let
\begin{align}
\phi_{(i,a), i}(y,x)&=
b_i(x)
\frac{\partial \log b_i(x)}{\partial \theta^{i,a}_y},&
\hat{\phi}_{(i,a), i}(x,y)&=
\frac{\partial \log b_i(x)}{\partial \eta^{i,a}_y}
\label{eq:b4}
\end{align}
for $y\in\mathcal{X}\setminus\{0\}$ and $x\in\mathcal{X}$.
Then, \eqref{eq:b4} satisfies~\eqref{eq:inv0}, i.e.,
$\sum_{x\in\mathcal{X}} \phi_{(i,a), i}(y,x)\hat{\phi}_{(i,a), i}(x,w) = \delta(y,w)$ for any $y, w\in\mathcal{X}\setminus\{\underline{0}\}$.
Similarly, let
\begin{align}
\phi_{(i,a), a}(y,x)&=
b_i(x)
\frac{\partial \log b_i(x)}{\partial \eta^{i,a}_y},&
\hat{\phi}_{(i,a), a}(x,y)&=
\frac{\partial \log b_i(x)}{\partial \theta^{i,a}_y}
\label{eq:b5}
\end{align}
for $y\in\mathcal{X}\setminus\{0\}$ and $x\in\mathcal{X}$.
Let $q\times q$ matrices $M$ and $\hat{M}$ be
\begin{align*}
M_{x,0} &:= b_i(x),&
M_{x,y} &:=
b_i(x)\frac{\partial \log b_i(x)}{\partial \theta^{i,a}_{y}}\\
\hat{M}_{0,x} &:= 1,&
\hat{M}_{y,x} &:=
\frac{\partial \log b_i(x)}{\partial \eta^{i,a}_{w}}.
\end{align*}
Then, it holds $\hat{M}M = I$ and hence $M\hat{M}=I$, i.e.,
\begin{align*}
&b_i(x)+
\sum_{y\in\mathcal{X}\setminus\{0\}}
b_i(x)\frac{\partial \log b_i(x)}{\partial \theta^{i,a}_{y}} \frac{\partial \log b_i(z)}{\partial \eta^{i,a}_{y}}
=\delta(x, z)\\
\Longleftrightarrow&
\sum_{y\in\mathcal{X}\setminus\{0\}}
\frac{\partial \log b_i(x)}{\partial \theta^{i,a}_{y}} \frac{\partial \log b_i(z)}{\partial \eta^{i,a}_{y}}
=
\delta(x,z)\frac1{b_i(z)}-1.
\end{align*}
The equation~\eqref{eq:contr} is obtained from the following three equalities
\begin{align*}
&\sum_{\bm{y}\in(\mathcal{X}\setminus\{0\})^{d_a(E')}}
\left\langle \prod_{i\in\partial a, (i,a)\in E'} \frac{\partial \log b_i(X_i)}{\partial \theta^{i,a}_{y_{i,a}}}\right\rangle_{b_a}\\
&\quad\cdot\prod_{i\in\partial a, (i,a)\in E'} \frac{\partial \log b_i(z_{(i,a),a})}{\partial \eta^{i,a}_{y_{i,a}}}
\\
&=
\left\langle \prod_{i\in\partial a, (i,a)\in E'} \frac{\delta\left(z_{(i,a),a}, X_i\right) - b_i(z_{(i,a),a})}{b_i(z_{(i,a),a})}\right\rangle_{b_a}
\end{align*}
\begin{align*}
&\sum_{\bm{y}\in(\mathcal{X}\setminus\{0\})^{d_i(E')}}
\left\langle \prod_{a\in\partial i, (i,a)\in E'} \frac{\partial \log b_i(X_i)}{\partial \eta^{i,a}_{y_{i,a}}}\right\rangle_{b_i}\\
&\quad\cdot\prod_{a\in\partial i, (i,a)\in E'} \frac{\partial \log b_i(z_{(i,a),i})}{\partial \theta^{i,a}_{y_{i,a}}}
\\
&=
\left\langle \prod_{a\in\partial i, (i,a)\in E'} \frac{\delta\left(z_{(i,a),i}, X_i\right) - b_i(z_{(i,a),i})}{b_i(z_{(i,a),i})}\right\rangle_{b_i}
\end{align*}
and
\begin{align*}
&\sum_{y_{i,a}\in\mathcal{X}\setminus\{0\}}
b_i(z_{(i,a),a})
\frac{\partial \log b_i(z_{(i,a),a})}{\partial \theta^{i,a}_{y_{i,a}}}\\
&\quad\cdot b_i(z_{(i,a),i})
\frac{\partial \log b_i(z_{(i,a),i})}{\partial \eta^{i,a}_{y_{i,a}}}\\
&=
\delta\left(z_{(i,a),a},z_{(i,a),i}\right)b_i(z_{(i,a),a})-b_i(z_{(i,a),a})b_i(z_{(i,a),i})\\
&=
\bigl\langle\left(\delta\left(z_{(i,a),a}, X_i\right)-b_i(z_{(i,a),a})\right)\\
&\qquad\cdot\left(\delta\left(z_{(i,a),i}, X_i\right)-b_i(z_{(i,a),i})\right)\bigr\rangle_{b_i}.
\end{align*}
\end{IEEEproof}
Lemma~\ref{lem:loopexp} shows that the weight of generalized loop in Theorem~\ref{thm:non-binary}
is equal to the weight of generalized loop obtained by Xiao and Zhou in Lemma~\ref{lem:xiao}~\cite{xiao2011partition}.

\bibliographystyle{IEEEtran}
\bibliography{IEEEabrv,biblio}

\end{document}